\documentclass[12pt,preprint2]{aastex}

\newcommand{\esoq}{ESO\,215-G?009}
\newcommand{\mcg}{MCG\,--04-02-003}
\newcommand{\atg}{ATCA\,J061608--574552}
\newcommand{\hi}{H\,{\sc i}}
\newcommand{\hii}{H\,{\sc ii}}
\newcommand{\mlr}{${\cal M}_{\rm HI}/L_{\rm B}$}

\newcommand{\gmr}{${\cal M}_{\rm gas}/{\cal M}_{\rm HI}$}

\newcommand{\msmbr}{${\cal M}_{*}/{\cal M}_{\rm bary}$}

\newcommand{\smlr}{${\cal M}_{*}/L_{\rm B}$}
\newcommand{\mls}{~${\cal M}_{\sun}/L_{\sun,{\rm B}}$}
\newcommand{\mmls}{${\cal M}_{\sun}/L_{\sun,{\rm B}}$}
\newcommand{\mB}{$m_{\rm B}$}

\newcommand{\MB}{$M_{\rm B}$}
\newcommand{\AB}{$A_{\rm B}$}

\newcommand{\LB}{$L_{\rm B}$}
\newcommand{\Lsun}{~$L_{\sun,{\rm B}}$}
\newcommand{\LLsun}{$L_{\sun,{\rm B}}$}
\newcommand{\Msun}{~${\cal M}_{\sun}$}

\newcommand{\FHI}{$F_{\rm HI}$}

\newcommand{\wxx}{$w_{\rm 20}$}
\newcommand{\MHI}{${\cal M}_{\rm HI}$}

\newcommand{\Mbary}{${\cal M}_{\rm bary}$}
\newcommand{\Mstar}{${\cal M}_{*}$}

\newcommand{\kms}{~km\,s$^{-1}$}
\newcommand{\kkms}{km\,s$^{-1}$}

\newcommand{\jjks}{Jy~km\,s$^{-1}$}

\newcommand{\vsys}{$v_{\rm sys}$}
\newcommand{\vlg}{$v_{\rm LG}$}
\newcommand{\pI}{Paper~{\rm I}}
\newcommand{\pII}{Paper~{\rm II}}

\slugcomment{}

\shorttitle{The Minimum Amount of Stars a Galaxy Will Form}
\shortauthors{Warren, Jerjen \& Koribalski}

\begin{document}
\title{The Minimum Amount of Stars a Galaxy Will Form}
\author{Bradley E. Warren\altaffilmark{1} and Helmut Jerjen}
\affil{Research School of Astronomy and Astrophysics, Australian National 
       University, Mount Stromlo Observatory, Cotter Road, Weston ACT 2611, 
       Australia}
\email{bwarren@physics.mcmaster.ca, jerjen@mso.anu.edu.au\\}
\and
\author{B\"arbel S. Koribalski}
\affil{Australia Telescope National Facility, CSIRO, PO Box 76, Epping NSW 1710, Australia}
\email{Baerbel.Koribalski@csiro.au}

\altaffiltext{1}{Affiliated with the Australia Telescope National Facility, CSIRO.  Now located at the Department of Physics and Astronomy, McMaster University, Hamilton, ON, Canada}

\begin{abstract}
We present an analysis of the atomic hydrogen and stellar properties of 38 late-type galaxies in the local Universe covering a wide range of \hi{} mass-to-light ratios (\mlr{}), stellar luminosities, and surface brightnesses.  Combining the results with those of four other well-studied dwarf galaxies known for their unusually large \hi{} contents, we identified an upper envelope for the \mlr{} as a function of galaxy luminosity.  This implies an empirical relation between the minimum amount of stars a galaxy will form and its initial baryonic mass. We also find that the star density systematically decreases with increasing \mlr{}, making the galaxies optically more elusive.

While the stellar mass of a galaxy seems to be only loosely connected to its baryonic mass, the latter quantity is strongly linked to the galaxy's dynamical mass as it is observed in the baryonic Tully-Fisher relation.  We find that dwarf irregular galaxies with generally high \mlr{}-ratios follow the same trend as defined by lower \mlr{} giant galaxies, but are underluminous for their rotation velocity to follow the trend in a stellar mass Tully-Fisher relation, suggesting that the baryonic mass of the dwarf galaxies is normal but they have failed to produced a sufficient amount of stars.

Finally, we present a three dimensional equivalent to the morphology-density relation which shows that high \mlr{} galaxies preferentially evolve and/or survive in low-density environments.  We conclude that an isolated galaxy with a shallow dark matter potential can retain a large portion of its baryonic matter in the form of gas, only producing a minimum quantity of stars necessary to maintain a stable gas disk.
\end{abstract}


\section{Introduction}
\label{sec:intro}

The formation and evolution of galaxies into the wide variety of stellar systems that we observe today, as well as their numbers and physical nature, is the source of much disagreement in theoretical and observational cosmology.  Traditionally to study the origins of galaxies observers have looked towards the farthest objects to examine those galaxies that are still in the process of formation, the regime of far-field cosmology (Hubble Deep Field \citealp{wil96}).  While this is an enormously important approach it faces obvious difficulties due to the very high redshifts of the targets.  Notably, only the most luminous systems can currently be examined, wavelength coverage is limited, and low angular resolution allows us to derive only global parameters.  Consequently we have less of a basis to compare very distant galaxies to nearby objects.

The expanding field of near field cosmology takes a different approach by looking at the evidence of galaxy evolution that remains in the local Universe.  This includes examination of stars within our own Galaxy for stellar systems that may have merged (the Sagittarius dwarf spheroidal [\citealp{iba94}], the Radial Velocity Experiment [RAVE, \citealp{ste06}] and Gaia [\citealp{per01}] surveys), searches (Sloan Digital Sky Survey, \citealp{yor00}) and detailed studies of satellites of the Milky Way and other galaxies in the Local Group (\citealp{gre04}; \citealp{tol99}; see \citealp{mat98} for a review), and exploration of galaxies in the Local Universe \citep{jer00,kar04,bou07,rej06} some of which have remained in an unevolved stage for most of the age of the Universe (such as DDO\,154 [\citealp{kru84}] or \esoq{} [\citealp{war04}]).  Dwarf galaxies (faint/low-mass galaxies) are particularly good probes of the cosmological past, being the baryonic manifestations of the dark matter building blocks.

In the last decade or so, a small but increasing number of unusual dwarf galaxies have been detected that hold a disproportionally large amount of atomic hydrogen gas (\hi{}) compared to their stellar content.  High quantities of \hi{} gas in these low luminosity systems suggest that star formation within them has been impaired or halted, has lacked stimulation, or has only recently begun.  Understanding why such extreme galaxies evolved since the epoch of reionisation, especially why they have produced only a small quantity of stars, is crucial for obtaining a more complete picture of galaxy formation and evolution.  Are they recently formed galaxies in which the extended \hi{} is only a temporary feature?  Alternatively, they could be genuine old galaxies that after an initial burst of star formation were prevented from further development or they had an extremely low star formation rate, just enough to keep the galaxy in a stable state.

The existence of galaxies in which the gas component dominates the baryonic matter, or even objects that are completely `dark' galaxies, has long been proposed, but observational difficulties have made them hard to find \citep{dis76}.  Although no galaxies entirely without stars have been found to date \citep{doy05,bek05}, up to half a dozen late-type dwarf galaxy examples are known in which the \hi{} gas appears to be the dominant form of detectable baryonic matter.  In the most extreme case, \esoq{} \citep[][ hereafter \pI{}]{war04}, there may be $\sim$20 times as much mass in atomic hydrogen as there is in stars.  Many of these objects, including \esoq{}, DDO\,154 \citep{car89,car98}, and NGC\,3741 \citep{beg05,gen07}, have \hi{} discs that extend 5 to 8 times the optical Holmberg radius.  Gas dominated galaxies are often cited as a solution to various ongoing cosmological problems, in particular the missing-satellites problem with Cold Dark Matter modeling \citep{ver02,dav06}, although other evidence suggests there are too few to account for the missing galaxies \citep[][ hereafter \pII{}]{tay05,war06}.

In our previous investigations (Papers~{\rm I} \& {\rm II}) we looked at the properties of nine individual dwarf galaxies in the local Universe chosen from the HIPASS {\it Bright Galaxy Catalog} \citep[][ hereafter HIPASS BGC]{kor04} because they initially appeared to have a high proportion of their baryonic mass in \hi{}, as measured by the \hi{} mass-to-light ratio (\mlr{}, which compares the \hi{} mass to the {\em B}-band luminosity).  We found that most of the original estimates for \mlr{} based on available but inaccurate photographic magnitudes from the literature were too high, and that genuine high \hi{} mass-to-light ratio ($>$5\mls{}) galaxies are rare in the local Universe.  The best remaining examples in our sample and in the literature generally have very extended \hi{} disks, are spatially isolated, and have normal baryonic content for their total masses, but are deficient in stars.  We now wish to understand how these trends hold up with a larger sample of low-mass galaxies across a broader range of \mlr{} and luminosity.

In this paper we carry out a statistical study of \hi{} and stellar properties of 38 nearby low-mass galaxies with a broad range of properties to find empirical input for the question how the baryon-to-light (or \hi{} gas mass-to-stellar mass) ratio changes as a function of observational quantities such as total baryonic mass, luminosity, surface brightness or environmental density.  Our results will characterize galaxies with unusually high \mlr{} and thus allow to predict where the theoretically proposed but still undiscovered baryonic dark galaxies are most likely to be found and how they may appear.

\section{Sample Selection and Observations}
\label{sec:sam}

Many past studies of gas-rich dwarf galaxies have selected samples based only on optical data (e.g. galaxies with low surface brightness or interesting morphology), or have come from samples of limited area or scope (e.g. from short scanning surveys or from studies of groups).  We have obtained multi-wavelength observations for 38 nearby field galaxies selected by their \hi{} properties \citep[see][]{war05}.  They were taken from the HIPASS BGC, which lists the 1000 \hi{}-brightest extragalactic sources (by \hi{} peak flux density) in the Southern hemisphere.  We chose mostly late-type galaxies with a wide range of \hi{} mass-to-light ratio, initially estimated by combining optical magnitudes from HyperLeda \citep[][ and references therein]{pat03} with the HIPASS BGC \hi{} flux densities and Galactic extinction data from \citet{sch98}.  The sample was limited to intrinsically faint objects (\MB{} $\gtrsim -16.5$~mag from the initial estimates of this value, although our new \MB{} estimates for these galaxies go up to $\sim -18$~mag), and selection favoured some objects with high initial estimates of \mlr{} but was otherwise random.  The sample contains all ten galaxies presented in \pII{} (including the newly discovered neighbor to ESO\,121-G020, \atg{}).

ATCA \hi{} line observations of our 38 sample galaxies were carried out between June 2002 and June 2003.  21 of the galaxies (those with $\delta < -30$\degr{}) were observed in different East-West configurations.  ESO\,121-G020, ESO\,428-G033, and ESO\,348-G009 were observed for $2 \times \sim12$ hours, \esoq{} was observed for $3 \times \sim$12 hours, while the remaining galaxies where observed in snapshot mode ($\sim1 - 2$ hours taken over a 12 hour period).  For the other 17 galaxies (with $\delta > -30$\degr{}) we used the compact hybrid arrays that include antennas on the Northern spur, resulting in rather large synthesized beams.  The galaxies \mcg{}, ESO\,505-G007, and IC\,4212 were observed for $\sim$10 hours, while were done in snapshot mode.  \hi{} snapshots that were initially taken for the galaxies with deeper observations (except \esoq{} and IC\,4212) were combined with the longer scans.  Details of the \hi{} observations for each galaxy are given in Table~\ref{tab:hiobs}.  The primary calibrator for all observations was PKS\,1934--638.

Data reduction and analysis were performed with the {\sc MIRIAD} package using standard procedures, with further analysis using {\sc AIPS}, {\sc GIPSY}, and {\sc KARMA}.  Channels affected by Galactic \hi{} emission were discarded where appropriate.  After continuum subtraction, the \hi{} data were Fourier-transformed using ``natural'' weighting and a channel width of 4\kms{}.  The data were cleaned and restored with the synthesized beam (the size of which is given in Table~\ref{tab:hiobs} for each galaxy).  Primary-beam correction was applied.  \hi{} distributions (0th moment) were obtained for all galaxies using cutoffs between 3 and 4$\sigma$.

Optical {\em B}-band CCD photometry was obtained with the Australian National University (ANU) 2.3m telescope at the Siding Spring Observatory as a series of 300\,s or 600\,s exposures during observing runs between April 2002 and February 2004, using the Nasmyth Imager (SITe $1124 \times 1024$ thinned CCD).  The imager has a circular field of view with a diameter of 6\farcm62 and a pixel size of 0\farcs59.  Table~\ref{tab:optobs} gives a summary of the observations taken for each galaxy in each band.  The columns are as follows: (2) broad-band (Cousins) filters used; (3) total observing time in each of the optical bands including the number of individual exposures; and (4) atmospheric seeing in the final optical images.  Most observations were taken at low airmass.  Twilight sky flat fields in all bands and bias images were obtained at the same time.  On each photometric night several \citet{lan92} standard stars were taken together with shallow 120\,s {\em B} images of the galaxy fields to perform the photometric calibration of the deeper images.  

Data reduction, photometric calibration, and analysis were carried out within IRAF using standard procedures.  After overscan subtraction, bias subtraction, and flatfielding, individual sets of images were registered and the sky level was subtracted. The images for each band were then combined into a single image (to increase the signal-to-noise, remove cosmic rays, etc.) and the photometric calibration applied.  Foreground stars were removed by replacing them with the surrounding sky so that only the galaxy remained.  Special care was taken to restore the light distribution under any stars superimposed onto the galaxies, e.g. using the mirror image from across the galaxies center.  For more details of this technique see \citet{jer03}.

Table~\ref{tab:rawprop} summarizes the details and main observational results for our 38 sample galaxies, including our new values of integrated \hi{} flux density (\FHI{}) and {\em B}-band apparent magnitude (\mB{}).  Our \hi{} flux densities are generally in good agreement with those from the HIPASS BGC, but most of the \mB{} values are brighter than the initial values from HyperLeda (see \pII{}).  With these new multiwavelength observations we can accurately measure quantities like \mlr{} that were previously only poorly estimated.  Table~\ref{tab:summary} lists properties derived for each galaxy from our observational results, including the \hi{} mass-to-light ratio.  With the exception of \esoq{}, distances were derived using the galaxy velocities relative to the barycenter of the Local Group.  Throughout this paper we adopt H$_0$ = 75\kms{}\,Mpc$^{-1}$.  In \esoq{}'s case, a tip of the Red Giant Branch distance measurement recently became available from \citet{kar07}.  So we could take advantage of this value as it represents a more reliable distance indicator for nearby galaxies where peculiar motions can be of comparable amplitude as the local Hubble flow.  We can now examine how the properties of these galaxies vary with \mlr{} to determine how some galaxies can maintain such a high proportion of neutral gas, while most known galaxies appear to have processed the majority of their gas into stars.

\section{An Upper Limit to the \hi{} Mass-to-Light Ratio}
\label{sec:slide}

\begin{figure*} 
  \plotone{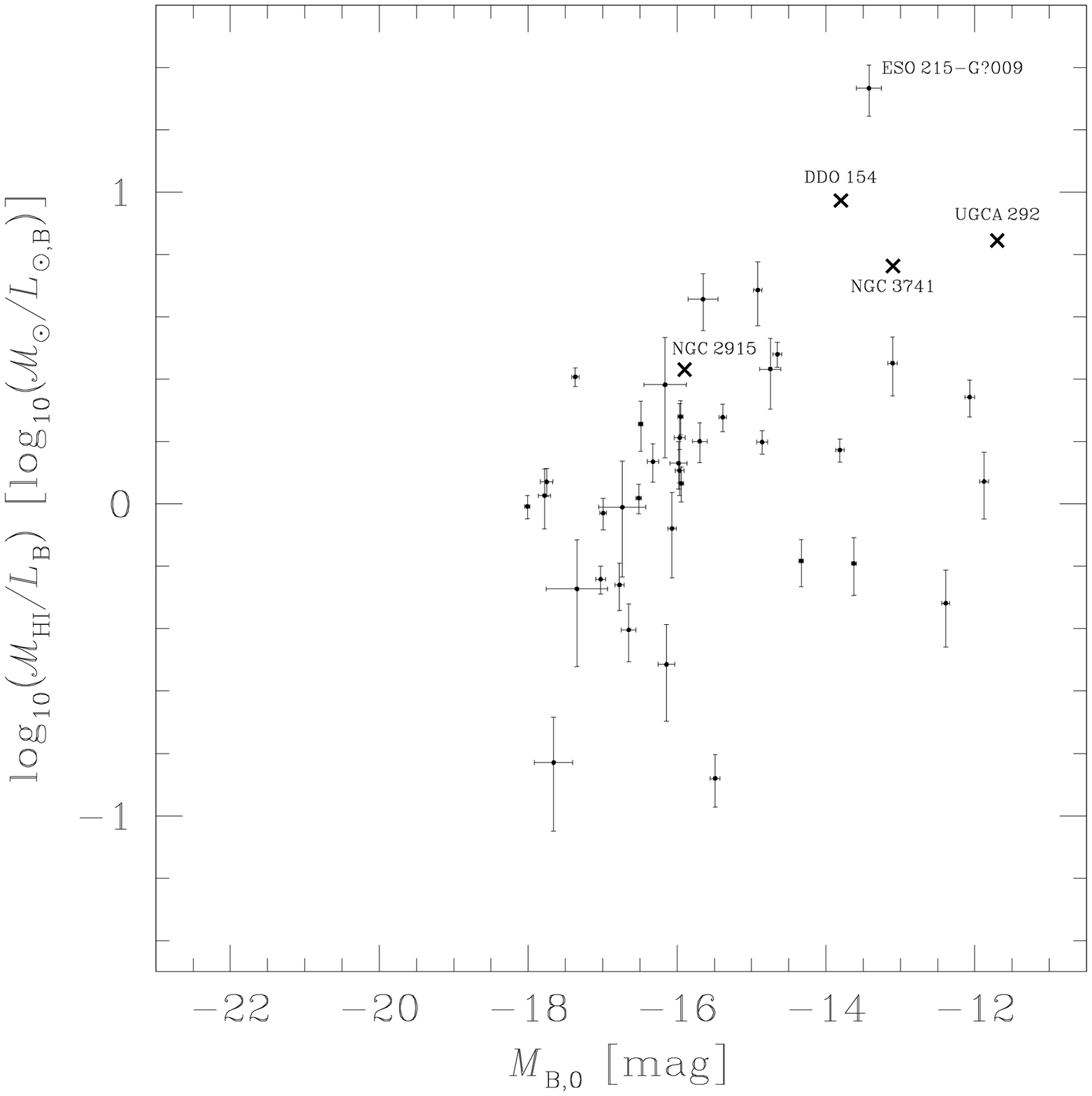}
\caption{\hi{} mass-to-light ratio versus absolute {\em B} magnitude for the 38 galaxies, including uncertainties as given in Table~\ref{tab:summary}.  Crosses with labels mark the position of four additional high \mlr{} galaxies taken from the literature.  Each of these four and \esoq{} are labelled.
\label{fig:mlmbt}}
\end{figure*}

The \hi{} mass-to-light ratio is a distance independent quantity that compares the mass of the atomic hydrogen gas to the luminosity (often the {\em B}-band luminosity).  It can therefore be considered a comparison of the relative amounts of the baryonic matter within a galaxy in the form of either gas or stars.  This ratio is well known to be higher on average with a greater range for galaxies with late-type morphologies \citep{rob94} and low luminosities (Papers~{\rm I} \& {\rm II}).  Fig.~\ref{fig:mlmbt} shows \mlr{} versus \MB{} (as a rough indicator of a galaxy's stellar mass) for the 38 galaxies in our sample (points with error bars).  

There are a few galaxies in our sample with high ratios (six with \mlr{} $\gtrsim$ 3\mls{}). \esoq{} in the top right corner clearly stands out. In order to have a more continuous coverage of the \mlr{} distribution we have included four well-studied galaxies from the literature known for their \hi{} extent.  Table~\ref{tab:other} summarizes the properties of these galaxies, DDO\,154 \citep{car89,car98}, NGC\,3741 \citep{beg05,gen07}, NGC\,2915 \citep{meu94,meu96}, and UGCA\,292 \citep{vzee01,you03}.  They are marked with crosses and labeled in Fig.~\ref{fig:mlmbt}.  Three of these galaxies (DDO\,154, NGC\,3741, and UGCA\,292) fit nicely into the gap between \esoq{} (also labeled) and the rest of our sample, while NGC\,2915 is comparable to the moderately high \mlr{} dwarf galaxies.  With the inclusion of these galaxies we have a sample with a good range in both \mlr{} (between about 0.15 and 20\mls{}) and luminosity (\MB{} between about -11.2 and -18.0 mag), with which we can make comparisons.  Importantly, we have a range in \mlr{} of about two orders of magnitude for galaxies of similar luminosities over most of the low luminosity range.

\begin{figure*} 
  \plotone{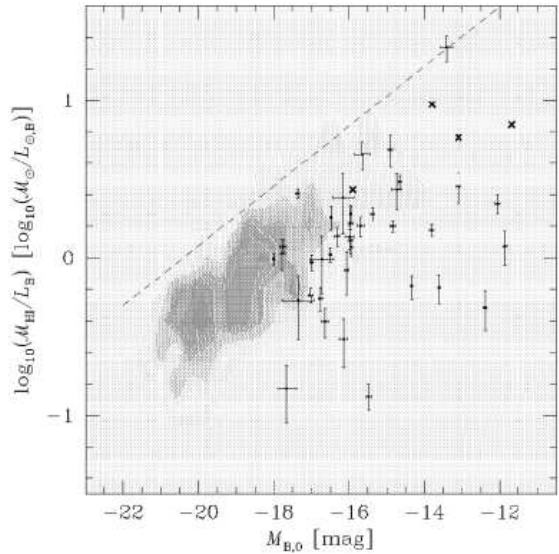}
\caption{Same data as in Fig.~\ref{fig:mlmbt} but complemented with the density distribution of 752 galaxies from the HIPASS BGC (in greyscale) that have HyperLeda magnitudes and that were not selected for our study.  The dashed line marks the locus of an upper envelope for the \hi{} mass-to-light ratio at a given luminosity.
\label{fig:mlmbtcon}}
\end{figure*}

The low-luminosity and low-\mlr{} region of Fig.~\ref{fig:mlmbt} (bottom right corner) is underpopulated in this \hi{}-selected sample because transition-type dwarf galaxies, such as studied by \citet{bea06} and \citet{bou05,bou07}, remain undetected by HIPASS.  Moreover, our study is focusing on dwarf galaxies (initially selected on the basis of their luminosity), and thus intrinsically bright galaxies (left side of Fig.~\ref{fig:mlmbt}) are naturally not present either.  Fig.~\ref{fig:mlmbtcon} shows the \mlr{} versus \MB{} as in Fig.~\ref{fig:mlmbt}, with the same axis scales, and again shows our sample and the four literature galaxies.  This time, in order to see the trend of the more luminous galaxies with \mlr{} we have included the density distribution of 752 HIPASS BGC galaxies (those that have HyperLeda apparent {\em B} magnitudes, and are not in our sample).  This sample covers a wide range of types and masses, but is dominated by luminous, gas-rich galaxies.  While the uncertainties in the optical measurements for these galaxies can be quite large (see \pII{} for a full discussion) the grey-scale representation of the data shall indicate the global trend of more luminous galaxies.

Galaxies with high \mlr{} values should be easily detectable in an \hi{} selected sample, especially if they are also intrinsically luminous.  So if such high \mlr{} galaxies exist they should be present to the upper left of Fig.~\ref{fig:mlmbtcon}.  However, brighter galaxies have only moderate ratios.  In fact the data are suggesting an upper limit for the \mlr{} at a given luminosity that increases systematically, with lower-luminosity galaxies able to reach higher \mlr{}.

In \pII{} we briefly mentioned that this trend may define an upper envelope for \mlr{}, a maximum amount of atomic hydrogen gas a galaxy of a particular luminosity can retain in the Universe today.  The dashed line in Fig.~\ref{fig:mlmbtcon} marks an empirically determined envelope for \mlr{} as a function of absolute magnitude (\MB{}).  This line is analytically best described by the equation:
\begin{equation}  
\log{\left( \frac{{\cal M}_{\rm HI}}{L_{\rm B}} \right)_{max}} = 0.19(M_{\rm B,0} + 20.4) ,
\label{eqn:envelope}
\end{equation} 
in units of $\log{({{\cal M}_{\sun}}/{L_{\sun,{\rm B}}})}$.  At a given absolute magnitude we would not expect the \mlr{} to exceed the value defined by this envelope.  This envelope could be further tested with the addition of accurate data for a sample of more luminous galaxies, and any fainter dwarfs galaxies with extreme \hi{} mass-to-light ratios if they exist.  It should be noted that the massive, low surface brightness galaxy Malin~1 \citep{imp89} potentially lies above this envelope, although calculations of the stellar and \hi{} properties of this galaxy are highly uncertain and upcoming results indicate that this object is possibly a member of a small galaxy group (Briggs, priv. comm.).  Using various results \citep{bot87,imp89,pic97,mat01} the \mlr{} for Malin~1 can range from $\sim$0.1 to 6\mls{}, which at its high luminosity of M$_B\approx-21.6\pm0.4$ could be either side of the envelope.

This upper envelope may reflect a true physical limitation on the \hi{} mass-to-light ratio for a galaxy of a certain initial mass.  As discussed in \pII{}, such an observation supports the idea that galaxies will produce a minimum quantity of stars from their initial gas cloud depending on the initial mass of the galaxy, as discussed theoretically by \citet{tay05}.  They found that without an internal radiation field the majority of the gas in the disk of a galaxy will become gravothermally unstable leading to the formation of H$_{2}$, even for galaxies with very low baryonic masses (down to $5\times10^{6}$\Msun{}).  They also found that the fraction of unstable gas decreases as the baryonic mass decreases, so that low-mass galaxies will stabilize with a lower fraction of gas converted to stars (assuming there are no other processes present to further drive star formation).  Hence galaxies such as \esoq{} that approach the upper envelope may represent objects that have formed only the minimum quantity of stars required to remain stable.

\begin{figure*} 
  \plotone{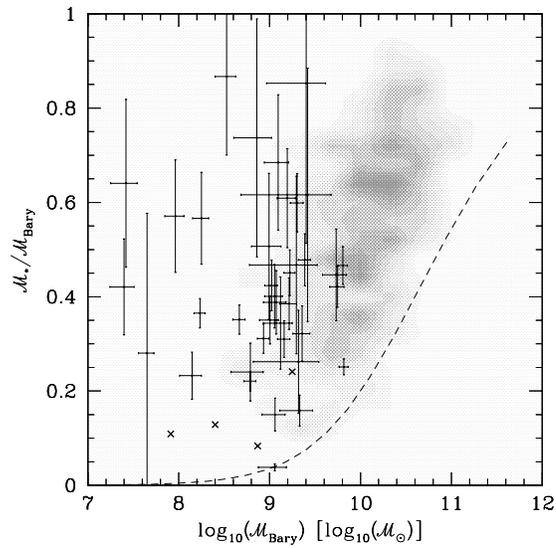}
\caption{The stellar mass fraction as a function of total baryonic mass for the 38 galaxies in our sample and the four literature galaxies (symbols as in Fig.~\ref{fig:mlmbt}).  The density distribution of the 752 galaxies in the HIPASS BGC with HyperLeda magnitudes (excluding sample galaxies) is again shown in greyscale.  The dashed line was derived from the upper envelope line marked in Fig.~\ref{fig:mlmbtcon}, and represents the minimum mass fraction of stars that a galaxy must form in order to remain gravothermally stable (see text).
\label{fig:mfract}}
\end{figure*}

We can use our upper envelope function to estimate the minimum fraction of the total baryonic mass that needs to be converted into stars for a galaxy of a given baryonic mass to remain gravothermally stable.  Fig.~\ref{fig:mfract} shows the mass fraction of baryons in the form of stars (${\cal M}_{*}/{\cal M}_{\rm bary}$) as a function of the total baryonic mass of the galaxy (stellar plus gas mass, \Mbary{}).  Our sample and the literature galaxies are marked as in Fig.~\ref{fig:mlmbt} and \ref{fig:mlmbtcon}, and the density distribution of the HIPASS BGC galaxies is included to illustrate the behavior of more luminous galaxies as in Fig.~\ref{fig:mlmbtcon}.  For this plot we have adopted a stellar mass-to-light ratio of \smlr{} = 1.2\mls{} to convert luminosity into stellar mass, and a total gas-to-\hi{} mass ratio of \gmr{} = 1.4 to convert \hi{} mass to the total gas mass (see \S~\ref{sec:tf} for further discussion of these values).  The dashed curve is a direct consequence of Equation~\ref{eqn:envelope}, derived by calculating ${\cal M}_{*}/{\cal M}_{\rm bary}$ and \Mbary{} from a series of \MB{} values.  It is best approximated analytically by the modified Gaussian function:

\begin{equation}  
\left(\frac{\mathcal{M}_*}{\mathcal{M}_{\rm Bary}}\right)_{\rm min}= \exp \left[-\left(\frac{15.7-\log(\mathcal{M_{\rm Bary}})}{5.2}\right)^{4.9}\right]
\label{eqn:envelerf}
\end{equation}

in the mass interval $7<\log(\mathcal{M}_{\rm Bary})<11.2$.  
This line can be interpreted as the minimum fraction of the total baryonic mass that must be converted into stars for a galaxy to become stable against large-scale star formation, assuming there are no other internal or external drivers of star formation.  The galaxies plotted in Fig.~\ref{fig:mfract} are all above this line, and the HIPASS BGC density distribution also falls well above.  Obtaining further accurate observations of galaxies that approach this envelope across the full range of luminosities/baryonic masses will help define how low the initial baryonic content of a galaxy has to be to allow a galaxy to form as a `dark' object.

\section{The Tully-Fisher Relationship}
\label{sec:tf}

The relationship between a galaxy's rotation velocity and its luminosity as explored by \citet{tul77} is a well established link between the properties of dark and luminous matter within disk galaxies, and is often used as a distance indicator \citep{tul77,pie88}.  However, at the faint end of the Tully-Fisher relationship, where the maximum rotation is low and the luminosities are faint, the \hi{} mass is a significant or even dominant fraction of the total baryonic mass.  Consequently, the relation is ill-defined there \citep{matt98,mcg00}.  That is why in recent years many authors have begun investigating a baryonic version of this relationship \citep{mcg00,bel01,gur04,pfe05,mcg05,geh06,der07}, where the \hi{} mass of the galaxy is included with the stellar mass.  Our sample of galaxies covers a wide range of baryonic composition from gas-dominated to star-dominated at the low-mass, low-rotation velocity end of this relation.  We therefore are in a position to check whether the high \mlr{} galaxies in our sample deviate significantly in Tully-Fisher plots from the rest of the sample and higher mass objects.

The rotation velocities of our sample galaxies were derived from newly measured ATCA \wxx{} values corrected for inclination using estimates of the stellar axis ratios from the optical images (Col (14) of Table~\ref{tab:rawprop}).  As the inclination for ESO\,428-G033 is highly uncertain (see \pII{}) it was excluded from this analysis.  The stellar mass was estimated from the {\em B}-band luminosity using a stellar mass-to-light ratio \smlr{} (as defined in the previous section), while the baryonic mass was estimated from the stellar mass plus the gas mass (from the \hi{} mass times a conversion factor \gmr{} to account for other components).

There is some difficulty in estimating both the stellar mass and the baryonic mass of a galaxy, in addition to both being dependent on an accurate distance measurement.  This is especially true in the dwarf regime where we often lack information on how metallicity and star formation affect our observed parameters, as well as the quantities of more difficult to detect components (such as dust and molecular gas).  The optical {\em B}-band, while probably still the most commonly observed wavelength for all galaxies, is a poor tracer of stellar mass compared to near-infrared imaging due to star formation and extinction effects \citep{bro92,rix93,rhe04}.  \citet{bel01} find that the stellar mass-to-light ratio can vary by a factor of as much as 7 at some optical wavelengths.  As we do not have direct information on \smlr{} for all our galaxies we have adopted a moderate ratio of 1.2\mls{} to convert the {\em B}-band luminosity to a stellar mass.

As the baryonic mass includes the stellar mass, it includes the uncertainties in that value too.  In addition, while we do have an accurate measure of the atomic hydrogen content of the galaxies, we do not know how much helium and molecular hydrogen gas are present.  We adopt the same conversion factor as \citet{mcg00} and \citet{geh06}, \gmr{} = 1.4, which takes into account helium and metals at cosmological abundances but not molecular hydrogen that is assumed to be less significant in dwarf galaxies \citep{tay98,ler05}.  However, little is know yet about the molecular gas content of such galaxies, and as their metalicity is often low using CO lines to trace H$_2$ is more difficult.  The only two galaxies in the sample for which $^{12}{\rm CO}(1-0)$ observations have been attempted previously, IC\,1574 \citep{tay98} and UGC\,06780 \citep{sau03}, remained undetected.

The baryonic Tully-Fisher fit of \citet{pfe05} suggests that the \gmr{} conversion factor could be as high as 2.98 (i.e. \hi{} is only one third of the total gas mass).  However, this result is based only on a slight reduction in the scatter of their fit to the relation (only a $\sim$0.013 reduction in RMS from \gmr{} = 1) rather than any observational/theoretical evidence.  The scatter in the Tully-Fisher plot can be affected by several other factors in addition to the gas-to-\hi{} mass ratio (such as star formation, metallicity, distance uncertainty, and the inclination correction to \wxx{}), and it seems unlikely that molecular gas would dominate over atomic gas to such a degree in low-mass galaxies, especially if the star formation is low.  The \citet{pfe05} result does show that there is a significant improvement in the scatter with the inclusion of the \hi{} gas.  Similarly, \citet{mcg05} tried to use various methods of determining \smlr{} to reduce the scatter.  While there is more theoretical basis to their method, it seems unlikely that an accurate understanding of these factors can be obtained from the baryonic Tully-Fisher relation at this stage without significantly reducing the other uncertainties that contribute to the scatter.

\begin{figure} 
  \plotone{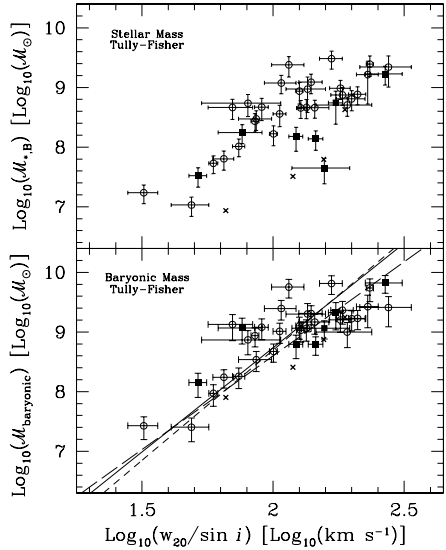}
\caption{The stellar mass (top) and baryonic mass (bottom) Tully-Fisher relations from our 38 sample objects (open circles and filled squares), and the four literature galaxies (crosses).  Stellar mass is calculated from the {\em B}-band magnitude.  The filled squares mark the galaxies in our sample with \mlr{} $> 2$\mls{}, with the open circles the remainder of the objects.  The three lines in the bottom panel marks the baryonic Tully-Fisher relations derived by \citet{bel01} (solid), \citet{geh06} (short dashed), and \citet{der07} (long dashed).
\label{fig:cbtf}}
\end{figure}

Fig.~\ref{fig:cbtf} shows both the stellar mass (top panel) and baryonic mass (bottom panel) Tully-Fisher relations for 37 of our sample galaxies (open circles and filled squares) and the four literature galaxies (crosses).  The baryonic version of this empirical relation is much tighter than the stellar mass only version, and the relation appears to hold down into the dwarf regime (\wxx{} $\sim$ 50\kms{}).  Unlike \citet{mcg00} and \citet{mcg05}, who used {\em I}-band data for the stellar mass, we do not see a distinct drop off in mass at the dwarf regime end in the stellar mass version of the relation, but merely a larger range in mass than is seen in the baryonic plot.  The \citet{mcg00} results contained a systematic error in their calculation of {\em I}-band magnitudes caused by a sign error in $V-I$, and \MHI{} was overestimated \citep{gur04}.  \citet{mcg05} only show this drop-off after converting \LB{} to \Mstar{}.

The black squares in both panels of Fig.~\ref{fig:cbtf} show the galaxies in our sample with \mlr{} greater than 2\mls{}.  In the case of the stellar-mass Tully-Fisher plot these galaxies and the four literature galaxies (crosses) in general sit at lower \Mstar{} values for their rotation velocity than the galaxies with lower \mlr{}.  \esoq{} in particular stands out in this panel (the square below the line of crosses, at $\log$(\wxx{}/$\sin i$) = 2.2, $\log$\Mstar{} = 7.6).  In the baryonic version though these galaxies have \Mbary{} values similar to the low-\mlr{} galaxies, and are generally within the scatter of the overall trend.  This reinforces the results of the study of \esoq{} (\pI{}), that high \hi{} mass-to-light ratio galaxies appear to have normal quantities of baryonic matter in them, but are deficient in stars.  This implies that these galaxies do not lack the ability to attract baryons, but are inefficient in forming stars.

The three lines going through the data of the baryonic Tully-Fisher plot (lower panel) are the least-squares fits derived by \citet{bel01} (solid line), \citet{geh06} (short dashed line), and \citet{der07} (long dashed line).  \citet{bel01} uses a sample of spiral galaxies in the Ursa Major Cluster with moderate to high masses obtained by \citet{ver97}.  \citet{geh06} compared their sample of very low-mass dwarf galaxies taken from the Sloan Digital Sky Survey \citep{bla05} to four other samples \citep{hay99,ver01,matt98,mcg00} to derive their relation.  \citet{der07} compared the \hi{} gas + stellar mass Tully-Fisher plot of a sample of dwarf elliptical (dE) galaxies to one derived from various literature sources \citep[including ][]{geh06,mcg00}, finding that the dE galaxies fall below the relation (potentially due to gas stripping in dense environments).  All three studies derive \smlr{} values from methods based on \citet{sal55} initial mass functions, and \citet{geh06} use \gmr{} = 1.4.  Our results for the baryonic Tully-Fisher relation are in good agreement with all three of these across the entire mass range that we cover.  We find that while the `classical' stellar-mass Tully-Fisher relation broadens significantly at the low-mass end, the baryonic relation appears to hold over the whole range of know galaxy masses.  Provided the significant uncertainties contributing to the scatter can be reduced to a minimum it could be a powerful distance indicator.  It may also eventually be useful for distinguishing other galaxy parameters such as gas composition, \smlr{}, and possibly the roll of gas stripping versus star formation in dense environments \citep{geh06,der07}.

\section{Stellar Density}
\label{sec:sbeff}

\begin{figure*} 
  \plotone{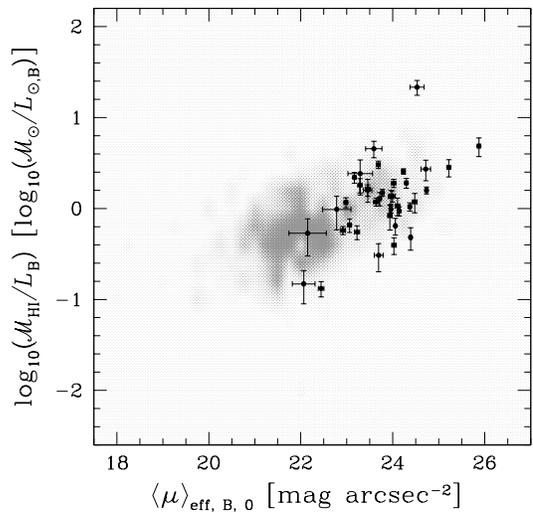}
\caption{\hi{} mass-to-light ratio versus the {\em B}-band effective surface brightness.  The solid points mark our 38 sample galaxies.  The greyscale shows the density distribution of 573 galaxies in the HIPASS BGC (those with effective surface brightness in HyperLeda and not in our sample.  Several galaxies with unusually large uncertainties in HyperLeda were also excluded).
\label{fig:sbeff}}
\end{figure*}

Fig.~\ref{fig:sbeff} shows \mlr{} versus the {\em B}-band mean effective surface brightness ($\langle \mu \rangle _{\rm eff,B}$).  The 38 galaxies in our sample are marked with the solid points with error bars, while the greyscale again shows the density distribution of galaxies from the HIPASS BGC (this parameter was only available for one of the literature galaxies, so they are not plotted).  We note that both quantities are distance independent.  There appears to be a strong correlation between these two values, with low-surface brightness (LSB) galaxies having higher \mlr{}'s, but it is important to understand the selection effects that may be at work here.

Similar to the way early-type galaxies were not seen in Fig.~\ref{fig:mlmbt}, low-\mlr{}, LSB galaxies are unlikely to be in the sample as they are both less likely to make the HIPASS BGC or to have been cataloged in the optical, so the lower region of the plot is unpopulated although we know such galaxies exist (i.e. dwarf ellipticals and other gas poor dwarfs).  Very compact galaxies that resemble stars, and therefore have very high surface brightness (HSB), are also unlikely to be cataloged, as was seen in the case of \atg{} (which has a star-like appearance, see \pII{}).  Whether these galaxies might also have high \hi{} mass-to-light ratios, and thus fall above the trend seen in Fig.~\ref{fig:sbeff} is a matter of speculation.  These galaxies with high stellar densities are likely to be experiencing a strong burst of star formation, which could mean that they have consumed or ejected much of their gas, or at least become bright enough to reduce \mlr{} to more typical levels.

If this was true, and compact high \mlr{} galaxies are not present, then the trend seen in Fig.~\ref{fig:sbeff} represents a limit to the \hi{} mass-to-light ratios for galaxies with a given stellar density, with only low density systems able to reach high \mlr{}'s.  The surface brightness is an indicator of the depth of the dark matter potential of the galaxy \citep{mo98}, with higher surface brightness galaxies having deeper potentials.  Therefore for a galaxy to end up with a high \mlr{} today it would need to initially have a shallow dark matter potential.

This picture supports well the results of \citet{deb96}, who looked at a sample of LSB galaxies in \hi{} drawn from the list of \citet{sch92}.  They found that their sample galaxies had higher \hi{} gas mass fractions than HSB galaxies for the same luminosity, an observation that appears independent of the photometric band in which the luminosity was measured.  When they looked at the relation between central surface brightness ($\mu_{\rm 0,B}$) and \mlr{} they found the same trend as we do for $\langle \mu \rangle _{\rm eff,B}$, with lower surface brightness galaxies having higher ratios.  We have chosen $\langle \mu \rangle _{\rm eff,B}$ over $\mu_{\rm 0,B}$ for our analysis as it is more representative of the overall stellar density of the galaxies given the difficulty in determining the morphological center of dwarf irregular galaxies.  These galaxies often have multiple \hii{} regions located close to their isophotal center (see for example the surface brightness profile of ESO\,505-G007 in \pII{}).

De Blok et al. (\citeyear{deb96}) also found a trend for LSB galaxies to have low \hi{} gas surface densities in comparison to their HSB counterparts.  This is consistent with the low gas surface density we have measured for \esoq{}.  If the dark matter potentials of these galaxies are shallow as implied by their low stellar and gas densities, then \citet{tay05} suggest that they may form into \citeauthor{too64}-stable disks.  This would cause the galaxy to have inefficient star formation (see \pI{}) and ultimately lead to the galaxy having a high \hi{} mass-to-light ratio, provided there were no other processes that disturbed the galaxy into more efficient star formation.

\section{The Effect of the Environment}
\label{sec:isolate}

The morphology-density relationship is now a well known correlation, generally characterized by a transition from greater numbers of \hi{}-rich late-type galaxies (spirals, dwarf irregulars) in low density environments to greater numbers of gas-poor early-type galaxies (giant and dwarf ellipticals) in dense environments \citep{oem74,dre80,bin87}.  It is particularly prominent in and around galaxy clusters, but has been seen to extend to several orders of magnitude lower in space density \citep{dre84}, into groups \citep{pos84} and the field \citep{bin90} where most of our sample galaxies originate.  Many of these classical morphology-density relations were derived from projected distances between the galaxies only (i.e. the line of sight separation was not taken into account).

Together with morphological appearance, the \hi{} content of a galaxy has been seen to correlate with the environmental density \citep{dre84,kar04}.  We have already shown in \pI{} that \esoq{} is a very isolated galaxy, being about 1.9~Mpc from the center of the nearest galaxy aggregate, the Centaurus A group, and $\sim$1.7~Mpc from NGC\,4945, which appears to be the nearest neighbor galaxy.  If there were a trend in our sample for the higher-\mlr{} dwarf galaxies to be generally more isolated than their low-\mlr{} counterparts this could suggest that the quantity of \hi{} a low-mass galaxy can retain, or in other words the efficiency of star formation in the galaxy, partially depends on the gravitational influence of neighboring galaxies even for the most isolated systems.

\citet{kar04} used a tidal index, $\Theta$, defined by \citet{kar99}, to quantify the tidal influence of neighboring galaxies on a large sample of galaxies (451) within 10~Mpc of the Milky Way (including most of our sample objects within this radius).  The index for a galaxy is based on the mass of the `main disturber' (the galaxy that appears to exert the largest tidal force) divided by the cube of the distance between the two, and is scaled such that objects with $\Theta < 0$ can be considered isolated.  \citet{beg05} have noted that the values of $\Theta$ given by \citet{kar04} for NGC\,3741, DDO\,154, and \esoq{} are all negative and all around $\Theta \sim -1$, suggesting a strong connection between high-\mlr{} galaxies and low-density environments.  Indeed, \citet{kar04} give negative indexes for the other two galaxies we have chosen from the literature as well.

In order to investigate how the environment could affect the \hi{} mass-to-light ratio we have looked at the neighbors of our sample galaxies and what influence they might have.  This is particularly difficult for several of these objects as they are very isolated, requiring accurate knowledge of all galaxies over a large sky area in the local volume to find the nearest and most influential neighbors.  This vast quantity of information is currently unavailable (although various studies are underway to approach this problem), so we are limited in what we can find for individual objects.  The NASA/IPAC Extragalactic Database (NED) has an option to search in a radius around a position (or a galaxy) for other objects in the database \citep[this was used by][to find the optical counterparts for the HIPASS BGC galaxies]{kor04}.  We took advantage of this search tool to find neighboring galaxies to the sample objects and extract some basic details on the neighbor, finding the galaxy with the smallest separation from each sample object (the `nearest' neighbor) using a rough estimate of the distance between the two objects, based on the sky separation and the recession velocity difference.  In this way we obtain a fully three-dimensional \mlr{}-density relation.

The NED search is limited to a maximum of 5\degr{} around the position chosen, which can restrict the search volume in the direction perpendicular to the line of sight for nearby galaxies quite severely.  For example, at 5.25~Mpc the NED search radius around \esoq{} for an object at the same distance from us is limited to $\sim460$~kpc.  For several galaxies, including \esoq{}, this radius was inadequate to find the nearest neighbor, all objects within 5\degr{} having redshifts well beyond the galaxy.  In these cases we had to resort to other search methods, such as searching at different positions around this radius or neighbors mentioned in literature sources \citep[several come from the analysis of][]{kar04}.  Finding neighbors for objects close to the Galactic Plane is also a problem as such galaxies could easily have been missed by the optical surveys that NED draws most heavily on.

The velocity separation is also a problem as both the sample galaxy and its neighbor could be affected by motions not related to the Hubble flow, especially in areas where they are subject to strong gravitational effects from clusters, groups, or even each other.  It is possible that ESO\,505-G007 is such a case, with the galaxy that contaminates its \hi{} measurements, ESO\,505-G008 (see \pII{}), potentially a close neighbor except for the large velocity difference that puts it much further away than other nearby objects.  Therefore, the galaxy found to be the nearest, and its estimated separation, only give an indication of the true environment of the sample galaxies as we may not have accurate distances between the galaxies.

Apart from looking at which galaxy is closest, we have also looked at which neighbor potentially has the greatest tidal influence on the galaxy.  Using the previously mentioned \citet{kar99} tidal index as a basis we determined the galaxy that is the main disturber, using the disturber's approximate luminosity as a rough indicator of mass (from the data available in NED), divided by the cube of the separation between the galaxy and the disturber.  In many cases the main disturber is the same galaxy as the nearest neighbor, especially where these galaxies are very close.  Table~\ref{tab:neigh} gives the details of both the nearest neighbors and the main disturber galaxies that we identified in NED for the 38 galaxies in our sample with new \mlr{} values, and the four literature galaxies.  For ESO\,368-G004, which is close to the Galactic Plane, we were unable to identify neighbors that are appropriate, so we have excluded it from this analysis

\begin{figure*} 
 \plotone{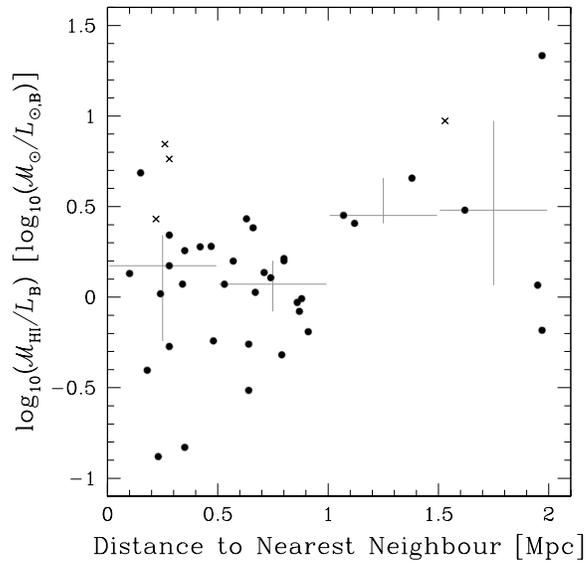}
\caption{\hi{} mass-to-light ratio versus the distance to the nearest neighboring galaxy for the 37 galaxies where we have both values (black points).  The black crosses (x) mark the positions of the four galaxies we have taken from the literature sources.  The horizontal grey lines mark the median \hi{} mass-to-light ratio for each 0.5 Mpc nearest neighboring distance bin, while the vertical grey lines mark the quartile range for each of those bins.
\label{fig:nn}}
\end{figure*}

Figure~\ref{fig:nn} shows the \hi{} mass-to-light ratio versus the spatial distance to the nearest neighbor for our 37 sample galaxies complemented with the data for the four literature galaxies.  Although there is a large scatter there is a trend for \mlr{} to increase as the galaxies become more isolated.  \esoq{} stands out at the top right, both isolated and with a high \mlr{}, with DDO\,154 between it and the other galaxies.  The majority of the sample has a neighbor within 1~Mpc, and almost all of these objects have low \mlr{}, all but one having a ratio less than 3\mls{}.  For the few more isolated galaxies there appears to be a greater spread in \mlr{}, but in general they have higher ratios.  In order to highlight any trends for \mlr{} with nearest neighbor distance we have put the data in 0.5 Mpc bins and calculated the median and quartile range of \mlr{} for each bin (shown as the grey lines in Fig.~\ref{fig:nn}).  The most isolated galaxies have a median \mlr{} of 3.0\mls{}, while the ones with the closest neighbors have a median \mlr{} of 1.5\mls{}.

\begin{figure*} 
 \plotone{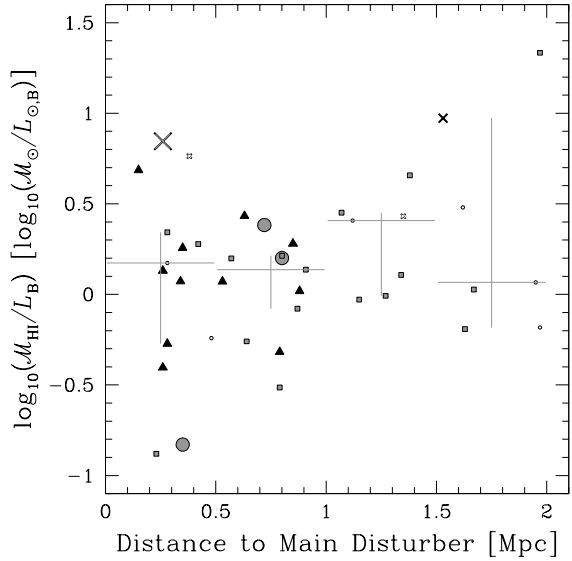}
\caption{\hi{} mass-to-light ratio versus the distance to the main disturber galaxy for the 37 galaxies where we have both values.  The crosses mark the positions of the four galaxies we have taken from the literature sources.  The sizes and shades indicate the approximate tidal effect of the disturber based on the \citet{kar99} tidal index.  Large grey circles/crosses indicate a strong effect ($\log_{10}(L_{\rm B, MD}/D^{3}_{\rm MD}) > 11$), medium black triangles/crosses are moderately affected ($11 > \log_{10}(L_{\rm B, MD}/D^{3}_{\rm MD}) > 10$), small grey squares/crosses are weakly affected ($10 > \log_{10}(L_{\rm B, MD}/D^{3}_{\rm MD}) > 9$), and the small open circles/crosses show those least affected by neighbors ($9 > \log_{10}(L_{\rm B, MD}/D^{3}_{\rm MD})$).  The horizontal grey lines mark the median \hi{} mass-to-light ratio for each 0.5 Mpc nearest neighboring distance bin, while the vertical grey lines mark the quartile range for each of those bins.
\label{fig:md}}
\end{figure*}

Fig.~\ref{fig:md} shows a similar plot but this time with the distances to the main disturber galaxies plotted for both our sample and the four literature galaxies.  The points and crosses are coded to indicate the approximate tidal effect of the disturber based on the \citet{kar99} tidal index, by dividing the approximate luminosity of the disturber by the cube of the separation between the galaxies.  Large grey circles/crosses indicate a strong effect ($\log_{10}(L_{\rm B, MD}/D^{3}_{\rm MD}) > 11$), medium black triangles/crosses a moderately effect ($11 > \log_{10}(L_{\rm B, MD}/D^{3}_{\rm MD}) > 10$), small grey squares/crosses a weak effect ($10 > \log_{10}(L_{\rm B, MD}/D^{3}_{\rm MD}) > 9$), and the small open circles/crosses show those least affected by neighbors ($9 > \log_{10}(L_{\rm B, MD}/D^{3}_{\rm MD})$).  Again we see that there is a general trend for more isolated galaxies to have higher \mlr{} than galaxies with close main disturbers.  There is also a slight trend for the galaxies that appear to be more strongly affected by tidal forces from their main disturber to have low to moderate \mlr{}.  Of the four literature galaxies only UGCA\,292 appears to have a neighbor with a strong tidal influence.  Again we have shown the median and quartile range of \mlr{} for each 0.5 Mpc bin in Fig.~\ref{fig:md}, although the trend is not as strong as for Fig.~\ref{fig:nn}.

The uncertainty in the separation values creates a large scatter in both plots.  The latter plot naturally has a larger uncertainty since in addition to the separation uncertainty we need to take into account uncertainties from the determination of the luminosity of the disturber.  However, both of these two plots suggest that spatial isolation is a crucial factor in the baryonic evolution of a dwarf galaxy.  The more isolated the galaxy the lower the star formation rate, and the lower the efficiency to convert \hi{} into stars.  Therefore, in order to find more high-\mlr{} galaxies, or even any true ``dark'' galaxies that may exist, we need to look at the low-galaxy density regions of the Local Universe.

\section{Discussion and Summary}
\label{sec:conc}

Why are there still so few high \mlr{} galaxies known today?  It has been more than two decades now since the fortuitous discovery of the unusual \hi{} properties of DDO\,154 \citep{kru84,car88}.  Since then very few galaxies similar to it have been found despite technological advances, wider and more extensive galaxy surveys (both blind and targeted), and a great increase in our knowledge of galaxies in general.  It is only in recent years that some galaxies with similar extreme \hi{} properties have been confirmed, such as UGCA\,292, NGC\,3741, and \esoq{}.  In addition, results from blind \hi{} surveys continue to provide strong evidence that there are no large numbers of genuine extragalactic \hi{} clouds without stars in the local universe \citep{rya02,doy05}.  It appears that there is more going on than the observational difficulties in detecting `dark' galaxies described by \citet{dis76}.  Dwarf galaxies in which the stellar content is only a small fraction of the total baryonic content appear to be genuinely rare objects, any internal and external influences being able to easily stimulate these objects produce more stars than the minimum required to remain stable.

At least two factors correlate with the \hi{} mass-to-light ratio and suggest possible reasons why they are so star deficient, and it appears that only a combination of these conditions can produce galaxies with very high ratios.  The stellar density of high-\mlr{} galaxies is lower than in other systems (\S~\ref{sec:sbeff}), which implies that the dark-matter potentials of these galaxies are shallow, making it more likely that the gas will not collapse to high enough densities for efficient star formation.  The highest-ratio galaxies have also formed in extreme isolation (\S~\ref{sec:isolate}), suggesting that tidal interactions can influence star formation even in relatively low density environments.  A galaxy needs to have a low initial mass and a shallow dark matter potential so that it can collapse into a \citeauthor{too64}-stable disk, but in most cases it will only remain this way if it continues in relative isolation.  
The minimum quantity of stars required for a galaxy of a certain mass to reach this point where it can form a stable disk appears to be well defined, as suggested by Fig.~\ref{fig:mfract}.

Dwarf galaxies with higher \mlr{} follow the same trend as lower \mlr{} galaxies in a baryonic Tully-Fisher plot, but are not massive enough for their rotation velocity on a stellar mass Tully-Fisher relation.  This suggests that the baryonic mass of these galaxies is normal but they have underproduced stars (\S~\ref{sec:tf}).  This is consistent with what we found in \pI{}, that when compared to the \citet{rob94} sample of typical dwarf galaxies the \hi{} mass and total mass of \esoq{} are at the lower ends of the respective quartile ranges, but the luminosity is more than an order of magnitude below this range.  It is also consistent with \citet{beg05}, who find that high-\mlr{} galaxies do not stand out in their fraction of total mass in the form of baryons, and in general that there is no trend of total-to-baryonic mass with luminosity.

Ultimately, the key to determining why high \mlr{} dwarf galaxies only form a small quantity of stars lies in reconstructing the star formation history.  \citet{gro07} looked at three recently discovered gas-rich dwarf galaxies in the Cen\,A group (1.4\mls{} $\le$ \mlr{} $\le$ 4.9\mls{}).  Morphologically these galaxies appear as low-surface brightness dwarf spheroidals, with little to no star formation activity. They obtained colour-magnitude diagrams (CMDs) from {\em V}- and {\em I}-band equivalent WFPC2 images on the Hubble Space Telescope, which at the distance of the Cen\,A group covers most of the Red Giant Branch (RGB).  Comparing the CMDs to evolutionary models they found that the stellar population of these galaxies is at least 2~Gyr old, and as much as 10~Gyr (the age-metallicity degeneracy prevents further refinement of this).
For their tip of the RGB distance for \esoq{} \citet{kar07} also obtained an {\em I} versus $V-I$ CMD for this galaxy, which shows a relatively broad RGB.  This may be at least in part due to the sky location of \esoq{} close to the Galactic Plane (high Galactic star contamination, uncertainties from dust extinction), but it may also indicate a range in stellar age (or metallicity).  Proper fits of evolutionary models would be needed to confirm this.  In any case there does appear to be a significant older stellar population in \esoq{}, and if we take this galaxy as typical of this class of object, high-\mlr{} galaxies are not newly formed galaxies experiencing their first burst of star formation.

\section*{Acknowledgments}

We are grateful for the assistance of Ken Freeman and Lister Staveley-Smith in this project, especially for their assistance with observations.  We would also like to thank Marilena Salvo and Gayandhi de Silva for their observing assistance.  Our thanks go to our referee, Uli Klein, for his helpful suggestions that have improved this article.  The 2.3-meter Telescope is run by the Australian National University as part of Research School of Astronomy and Astrophysics.  The Australia Telescope Compact Array and the Parkes Radio Telescope are part of the Australia Telescope that is funded by the Commonwealth of Australia for operation as a National Facility managed by CSIRO.  This research has made use of the NASA/IPAC Extragalactic Database (NED), which is operated by the Jet Propulsion Laboratory, California Institute of Technology, under contract with the National Aeronautics and Space Administration.

\clearpage


\begin{deluxetable}{lcrrl} 
\tabletypesize{\scriptsize}
\tablecaption{Summary of \hi{} Observations for each Galaxy. 
\label{tab:hiobs}}
\tablewidth{0pt}
\tablehead{
   \colhead{Name} & \colhead{Arrays} & \colhead{Time}    & \colhead{Central Freq.} & \colhead{Phase Cal.} \\
   \colhead{}     & \colhead{}       & \colhead{(hours)} & \colhead{(MHz)}         & \colhead{}           \\
   \colhead{(1)}  & \colhead{(2)}    & \colhead{(3)}     & \colhead{(4)}           & \colhead{(5)}        }
\startdata
 ESO\,349-G031     & EW352  &  $\sim1.9$ & 1417 & PKS\,0008--421 \\

 \mcg{}            & H75B   &  $\sim1.5$ & 1417 & PKS\,0023--263 \\
                   & H168B  &  $\sim8.6$ & 1417 & PKS\,0023--263 \\

 ESO\,473-G024     & H75B   &  $\sim1.5$ & 1417 & PKS\,0023--263 \\
                   & H168B  &  $\sim1.0$ & 1417 & PKS\,0023--263 \\

 IC\,1574          & H75B   &  $\sim1.5$ & 1417 & PKS\,0023--263 \\
                   & H168B  &  $\sim1.0$ & 1417 & PKS\,0023--263 \\

 UGCA\,015         & H75B   &  $\sim1.3$ & 1417 & PKS\,0023--263 \\
                   & H168B  &  $\sim1.0$ & 1417 & PKS\,0023--263 \\

 ESO\,085-G047     & EW352  &  $\sim3.0$ & 1414 & PKS\,0537--441 \\

 ESO\,120-G021     & EW352  &  $\sim2.2$ & 1414 & PKS\,0537--441 \\

 ESO\,425-G001     & EW352  &  $\sim0.8$ & 1414 & PKS\,0614--349 \\

 UGCA\,120         & H75B   &  $\sim2.2$ & 1417 & PKS\,0704--231 \\

 ESO\,121-G020\tablenotemark{a}     & 750D   & $\sim10.5$ & 1417 & PKS\,0407--658 \\
                   & 1.5B   & $\sim10.9$ & 1417 & PKS\,0407--658 \\
                   & EW352  &  $\sim2.4$ & 1416 & PKS\,0537--441 \\

 \atg{}\tablenotemark{a}            & 750D   & $\sim10.5$ & 1417 & PKS\,0407--658 \\
                   & 1.5B   & $\sim10.9$ & 1417 & PKS\,0407--658 \\
                   & EW352  &  $\sim2.4$ & 1416 & PKS\,0537--441 \\

 WHI\,B0619--07    & H75B   &  $\sim1.8$ & 1417 & PKS\,0704--231 \\

 ESO\,490-G017     & H75B   &  $\sim2.0$ & 1417 & PKS\,0704--231 \\

 ESO\,255-G019     & EW352  &  $\sim0.8$ & 1414 & PKS\,0614--349 \\
                   & EW367B &  $\sim0.8$ & 1414 & PKS\,0614--349 \\

 ESO\,207-G007     & EW352  &  $\sim2.0$ & 1416 & PKS\,0537--441 \\

 ESO\,207-G022     & EW352  &  $\sim2.0$ & 1416 & PKS\,0537--441 \\

 ESO\,428-G033     & 750D   & $\sim10.6$ & 1412 & PKS\,0614--349 \\
                   & 1.5B   & $\sim10.3$ & 1412 & PKS\,0614--349 \\
                   & EW352  &  $\sim1.1$ & 1414 & PKS\,0614--349 \\
                   & EW367B &  $\sim0.8$ & 1414 & PKS\,0614--349 \\

 ESO\,257-G?017    & EW352  &  $\sim0.8$ & 1414 & PKS\,0614--349 \\
                   & EW367B &  $\sim0.8$ & 1414 & PKS\,0614--349 \\

 ESO\,368-G004     & EW352  &  $\sim1.0$ & 1414 & PKS\,0614--349 \\
                   & EW367B &  $\sim0.8$ & 1414 & PKS\,0614--349 \\

 PGC\,023156       & EW352  &  $\sim1.0$ & 1414 & PKS\,0614--349 \\
                   & EW367B &  $\sim0.7$ & 1414 & PKS\,0614--349 \\

 ESO\,164-G?010    & EW352  &  $\sim2.7$ & 1416 & PKS\,0537--441 \\

 \esoq{}           & EW352  & $\sim11.6$ & 1417 & PKS\,1215--457 \\
                   & 750A   & $\sim10.6$ & 1417 & PKS\,1215--457 \\
                   & 6A     & $\sim11.5$ & 1417 & PKS\,1215--457 \\

 CGCG\,012-022     & H75B   &  $\sim1.0$ & 1412 & PKS\,1127--145 \\

 UGC\,06780        & H75B   &  $\sim1.0$ & 1412 & PKS\,1127--145 \\

 ESO\,572-G009     & H75B   &  $\sim1.0$ & 1412 & PKS\,1127--145 \\

 ESO\,505-G007     & H75B   &  $\sim1.0$ & 1412 & PKS\,1127--145 \\
                   & H168B  &  $\sim1.9$ & 1412 & PKS\,1127--145 \\
                   & H75B   &  $\sim9.2$ & 1412 & PKS\,1151--348 \\

 ESO\,572-G052     & H75B   &  $\sim1.0$ & 1412 & PKS\,1127--145 \\

 UGCA\,289         & H75B   &  $\sim1.4$ & 1415 & PKS\,1245--197 \\

 UGCA\,307         & H75B   &  $\sim1.3$ & 1415 & PKS\,1245--197 \\

 UGCA\,312         & H75B   &  $\sim1.3$ & 1415 & PKS\,1245--197 \\

 UGCA\,322         & H75B   &  $\sim1.4$ & 1415 & PKS\,1245--197 \\

 IC\,4212          & H75B   &  $\sim8.8$ & 1413 & PKS\,1308--220 \\

 IC\,4824          & EW352  &  $\sim0.7$ & 1416 & PKS\,1934--638 \\
                   & EW367B &  $\sim1.8$ & 1416 & PKS\,1934--638 \\

 ESO\,141-G042     & EW352  &  $\sim0.7$ & 1416 & PKS\,1934--638 \\
                   & EW367B &  $\sim1.8$ & 1416 & PKS\,1934--638 \\

 IC\,4870          & EW352  &  $\sim0.7$ & 1416 & PKS\,1934--638 \\
                   & EW367B &  $\sim1.8$ & 1416 & PKS\,1934--638 \\

 IC\,4951          & EW352  &  $\sim0.7$ & 1416 & PKS\,1934--638 \\
                   & EW367B &  $\sim1.8$ & 1416 & PKS\,1934--638 \\

 ESO\,348-G009     & 750D   & $\sim10.6$ & 1417 & PKS\,0008--421 \\
                   & 1.5B   &  $\sim9.8$ & 1417 & PKS\,0008--421 \\
                   & EW352  &  $\sim1.7$ & 1417 & PKS\,0008--421 \\

 ESO\,149-G003     & EW352  &  $\sim2.2$ & 1417 & PKS\,0008--421 \\
\enddata
\tablenotetext{a}{ESO\,121-G020 and \atg{} were observed in the same field}

\begin{flushleft}
(1) Galaxy name. \\
(2) Array configurations on the ATCA on which each galaxy was observed in the radio. \\
(3) Approximate time on source for each array. \\
(4) Central frequency of the radio \hi{} line observations for each galaxies. \\
(5) Radio phase calibration source used throughout the observations of each galaxy.
\vspace{9cm}
\end{flushleft}
\end{deluxetable}

\clearpage

\begin{deluxetable}{lcc} 
\tabletypesize{\scriptsize}
\tablecaption{Observing Details for {\em B}-band Photometry of Our Sample. 
\label{tab:optobs}}
\tablewidth{0pt}
\tablehead{
   \colhead{Name} & \colhead{Exposure Time} & \colhead{Seeing}   \\
   \colhead{}     & \colhead{(seconds)}     & \colhead{(arcsec)} \\
   \colhead{(1)}  & \colhead{(2)}           & \colhead{(3)}      }
\startdata
 ESO\,349-G031     & 3000 ($10\times300$) & 1\farcs9 \\

 \mcg{}            & 3000 ($10\times300$) & 2\farcs2 \\

 ESO\,473-G024     & 3000 ($10\times300$) & 2\farcs5 \\

 IC\,1574          & 3000 ($10\times300$) & 2\farcs0 \\

 UGCA\,015         & 3000 ($10\times300$) & 2\farcs2 \\

 ESO\,085-G047     & 3000 \phantom{1}($5\times600$)  & 2\farcs7 \\

 ESO\,120-G021     & 3000 ($10\times300$) & 2\farcs2 \\

 ESO\,425-G001     & 3000 \phantom{1}($5\times600$)  & 2\farcs9 \\

 UGCA\,120         & 3000 \phantom{1}($5\times600$)  & 2\farcs2 \\

 ESO\,121-G020\tablenotemark{a}     & 3000 ($10\times300$) & 2\farcs1 \\

 \atg{}\tablenotemark{a}            & 3000 ($10\times300$) & 2\farcs1 \\

 WHI\,B0619--07    & 3000 ($10\times300$) & 1\farcs5 \\

 ESO\,490-G017     & 3000 \phantom{1}($5\times600$)  & 2\farcs0 \\

 ESO\,255-G019     & 3000 \phantom{1}($5\times600$)  & 1\farcs9 \\

 ESO\,207-G007     & 3000 \phantom{1}($5\times600$)  & 2\farcs0 \\

 ESO\,207-G022     & 3000 \phantom{1}($5\times600$)  & 1\farcs9 \\

 ESO\,428-G033     & 3000 \phantom{1}($5\times600$)  & 2\farcs2 \\

 ESO\,257-G?017    & 3000 \phantom{1}($5\times600$)  & 1\farcs8 \\

 ESO\,368-G004     & 3000 \phantom{1}($5\times600$)  & 1\farcs9 \\

 PGC\,023156       & 3000 \phantom{1}($5\times600$)  & 1\farcs8 \\

 ESO\,164-G?010    & 3000 \phantom{1}($5\times600$)  & 2\farcs0 \\

 \esoq{}           & 3000 ($10\times300$) & 1\farcs9 \\

 CGCG\,012-022     & 3000 \phantom{1}($5\times600$)  & 1\farcs9 \\

 UGC\,06780        & 3000 \phantom{1}($5\times600$)  & 2\farcs2 \\

 ESO\,572-G009     & 1800 \phantom{1}($3\times600$)  & 1\farcs9 \\

 ESO\,505-G007     & 3000 ($10\times300$) & 2\farcs0 \\

 ESO\,572-G052     & 3000 \phantom{1}($5\times600$)  & 1\farcs8 \\

 UGCA\,289         & 3000 \phantom{1}($5\times600$)  & 2\farcs3 \\

 UGCA\,307         & 3000 \phantom{1}($5\times600$)  & 2\farcs0 \\

 UGCA\,312         & 3000 \phantom{1}($5\times600$)  & 2\farcs1 \\

 UGCA\,322         & 3000 \phantom{1}($5\times600$)  & 2\farcs5 \\

 IC\,4212          & 3000 \phantom{1}($5\times600$)  & 2\farcs9 \\

 IC\,4824          & 3000 ($10\times300$) & 1\farcs7 \\

 ESO\,141-G042     & 3000 ($10\times300$) & 1\farcs7 \\

 IC\,4870          & 3000 ($10\times300$) & 2\farcs1 \\

 IC\,4951          & 3000 ($10\times300$) & 2\farcs8 \\

 ESO\,348-G009     & 3000 ($10\times300$) & 1\farcs7 \\

 ESO\,149-G003     & 3000 ($10\times300$) & 2\farcs2 \\
\enddata
\tablenotetext{a}{ESO\,121-G020 and \atg{} were observed in the same field}

\begin{flushleft}
(1) Galaxy name. \\
(2) Total observing time in each of the optical bands including the number of individual exposures. \\
(3) Seeing in the final optical images in each band. \\
\vspace{9cm}
\end{flushleft}
\end{deluxetable}

\begin{deluxetable}{llcccccccccccc} 
\rotate
\tabletypesize{\scriptsize}
\setlength{\tabcolsep}{0.02in}
\tablecaption{Summary of the Basic Properties for the 38 Sample Galaxies.
\label{tab:rawprop}}
\tablewidth{0pt}
\tablehead{ & & & & & & \multicolumn{4}{c}{ATCA} & & \multicolumn{3}{c}{2.3m Telescope} \\ \cline{7-10} \cline{12-14}

\colhead{Name} & \colhead{HIPASS} & \colhead{$\alpha$} & \colhead{$\delta$} & \colhead{$l$} & \colhead{$b$} & \colhead{\FHI{}} & \colhead{\vsys{}} & \colhead{\vlg{}} & \colhead{\wxx{}} & \colhead{\AB{}} & \colhead{\mB{}} & \colhead{$\langle \mu \rangle _{\rm eff,B}$} & \colhead{$i$} \\

 \colhead{} & \colhead{Name} & \colhead{(J2000.0)} & \colhead{(J2000.0)} & \colhead{} & \colhead{} & \colhead{\jjks{}} & \colhead{\kkms{}} & \colhead{\kkms{}} & \colhead{\kkms{}} & \colhead{(mag)} & \colhead{mag} & \colhead{(mag arcsec$^{-2}$)} & \colhead{deg} \\
 
 \colhead{(1)} & \colhead{(2)} & \colhead{(3)} & \colhead{(4)} & \colhead{(5)} & \colhead{(6)} & \colhead{(7)} & \colhead{(8)} & \colhead{(9)} & \colhead{(10)} & \colhead{(11)} & \colhead{(12)} & \colhead{(13)} & \colhead{(14)}}
\startdata
 ESO\,349-G031     & J0008--34 & $00^{\rm h}\,08^{\rm m}$\,13\fs0 & --34\degr\,34\arcmin\,42\arcsec{} & 351\fdg6 & --78\fdg1 & $ 4.3 \pm 0.8$ & $ 227 \pm 2$ & $ 218 \pm 2$ & $ 40 \pm 4$ & $0.05 \pm 0.01$ & $15.71 \pm 0.06$ & $24.53 \pm 0.05$ & 55 \\

 \mcg{}            & J0019--22 & $00^{\rm h}\,19^{\rm m}$\,11\fs4 & --22\degr\,40\arcmin\,06\arcsec{} &  62\fdg5 & --81\fdg4 & $16.2 \pm 0.6$ & $ 670 \pm 2$ & $ 710 \pm 2$ & $126 \pm 2$ & $0.08 \pm 0.01$ & $15.32 \pm 0.06$ & $23.77 \pm 0.04$ & 60 \\

 ESO\,473-G024     & J0031--22 & $00^{\rm h}\,31^{\rm m}$\,22\fs5 & --22\degr\,45\arcmin\,57\arcsec{} &  75\fdg7 & --83\fdg7 & $ 5.7 \pm 0.9$ & $ 542 \pm 3$ & $ 574 \pm 3$ & $ 50 \pm 3$ & $0.08 \pm 0.01$ & $16.38 \pm 0.06$ & $25.30 \pm 0.04$ & 75 \\

 IC\,1574          & J0043--22 & $00^{\rm h}\,43^{\rm m}$\,03\fs0 & --22\degr\,14\arcmin\,49\arcsec{} & 101\fdg2 & --84\fdg7 & $ 5.0 \pm 0.9$ & $ 361 \pm 1$ & $ 388 \pm 1$ & $ 59 \pm 2$ & $0.07 \pm 0.01$ & $14.90 \pm 0.03$ & $24.12 \pm 0.02$ & 85 \\

 UGCA\,015         & J0049--20 & $00^{\rm h}\,49^{\rm m}$\,49\fs0 & --21\degr\,00\arcmin\,54\arcsec{} & 118\fdg9 & --83\fdg9 & $ 2.6 \pm 0.6$ & $ 301 \pm 2$ & $ 329 \pm 2$ & $ 32 \pm 4$ & $0.07 \pm 0.01$ & $15.30 \pm 0.05$ & $24.46 \pm 0.02$ & 85 \\

 ESO\,085-G047     & J0507--63 & $05^{\rm h}\,07^{\rm m}$\,43\fs0 & --62\degr\,59\arcmin\,30\arcsec{} & 272\fdg8 & --35\fdg7 & $14.7 \pm 1.1$ & $1445 \pm 2$ & $1202 \pm 2$ & $ 76 \pm 2$ & $0.11 \pm 0.02$ & $14.18 \pm 0.04$ & $24.25 \pm 0.02$ & 45 \\

 ESO\,120-G021     & J0553--59 & $05^{\rm h}\,53^{\rm m}$\,14\fs0 & --59\degr\,03\arcmin\,58\arcsec{} & 267\fdg7 & --30\fdg4 & $10.6 \pm 1.2$ & $1299 \pm 2$ & $1040 \pm 2$ & $132 \pm 4$ & $0.20 \pm 0.03$ & $14.96 \pm 0.05$ & $23.91 \pm 0.02$ & 80 \\

 ESO\,425-G001     & J0600--31 & $06^{\rm h}\,00^{\rm m}$\,10\fs0 & --31\degr\,47\arcmin\,16\arcsec{} & 237\fdg8 & --24\fdg1 & $ 6.7 \pm 1.7$ & $1341 \pm 2$ & $1109 \pm 2$ & $110 \pm 4$ & $0.17 \pm 0.03$ & $14.97 \pm 0.05$ & $24.11 \pm 0.04$ & 35 \\

 UGCA\,120         & J0611--21 & $06^{\rm h}\,11^{\rm m}$\,16\fs0 & --21\degr\,35\arcmin\,56\arcsec{} & 228\fdg4 & --18\fdg2 & $33.4 \pm 1.4$ & $ 859 \pm 1$ & $ 646 \pm 1$ & $107 \pm 2$ & $0.33 \pm 0.05$ & $12.98 \pm 0.04$ & $23.25 \pm 0.02$ & 50 \\

 ESO\,121-G020     & J0615--57\tablenotemark{a} & $06^{\rm h}\,15^{\rm m}$\,54\fs2 & --57\degr\,43\arcmin\,32\arcsec{} & 266\fdg5 & --27\fdg3 & $ 9.1 \pm 0.3$ & $ 583 \pm 2$ & $ 317 \pm 2$ & $ 61 \pm 4$ & $0.17 \pm 0.03$ & $15.27 \pm 0.05$ & $23.95 \pm 0.02$ & 70 \\

 \atg{}            & J0615--57\tablenotemark{a} & $06^{\rm h}\,16^{\rm m}$\,08\fs4 & --57\degr\,45\arcmin\,52\arcsec{} & 266\fdg6 & --27\fdg3 & $ 2.7 \pm 0.2$ & $ 554 \pm 4$ & $ 288 \pm 4$ & $ 56 \pm 8$ & $0.17 \pm 0.03$ & $17.01 \pm 0.06$ & $23.34 \pm 0.02$ & 25 \\
 
 WHI\,B0619--07\tablenotemark{b}    & J0622--07 & $06^{\rm h}\,22^{\rm m}$\,12\fs0 & --07\degr\,50\arcmin\,21\arcsec{} & 216\fdg7 & --10\fdg0 & $52.0 \pm 3.0$ & $ 759 \pm 1$ & $ 582 \pm 1$ & $176 \pm 2$ & $2.65 \pm 0.42$ & $14.63 \pm 0.06$ & $24.69 \pm 0.03$ & 50 \\

 ESO\,490-G017     & J0638--26 & $06^{\rm h}\,37^{\rm m}$\,57\fs0 & --26\degr\,00\arcmin\,01\arcsec{} & 235\fdg1 & --14\fdg3 & $ 7.7 \pm 1.0$ & $ 499 \pm 2$ & $ 261 \pm 2$ & $ 61 \pm 2$ & $0.33 \pm 0.05$ & $12.98 \pm 0.04$ & $22.78 \pm 0.03$ & 45 \\

 ESO\,255-G019     & J0645--47 & $06^{\rm h}\,45^{\rm m}$\,48\fs0 & --47\degr\,31\arcmin\,50\arcsec{} & 256\fdg8 & --20\fdg6 & $28.0 \pm 2.0$ & $1050 \pm 1$ & $ 777 \pm 1$ & $141 \pm 2$ & $0.36 \pm 0.05$ & $14.14 \pm 0.05$ & $24.35 \pm 0.03$ & 45 \\

 ESO\,207-G007     & J0650--52 & $06^{\rm h}\,50^{\rm m}$\,39\fs0 & --52\degr\,08\arcmin\,30\arcsec{} & 261\fdg8 & --21\fdg2 & $16.4 \pm 1.9$ & $1070 \pm 1$ & $ 793 \pm 1$ & $148 \pm 4$ & $0.33 \pm 0.05$ & $13.71 \pm 0.03$ & $23.56 \pm 0.02$ & 55 \\

 ESO\,207-G022     & J0709--51 & $07^{\rm h}\,09^{\rm m}$\,10\fs0 & --51\degr\,28\arcmin\,01\arcsec{} & 262\fdg1 & --18\fdg3 & $ 5.4 \pm 1.3$ & $1059 \pm 1$ & $ 777 \pm 1$ & $ 46 \pm 6$ & $0.56 \pm 0.09$ & $14.52 \pm 0.06$ & $24.27 \pm 0.03$ & 35 \\

 ESO\,428-G033    & J0725--30B & $07^{\rm h}\,25^{\rm m}$\,49\fs5 & --30\degr\,55\arcmin\,09\arcsec{} & 244\fdg2 &  --6\fdg9 & $14.5 \pm 0.3$ & $1728 \pm 2$ & $1460 \pm 2$ & $110 \pm 2$ & $1.10 \pm 0.18$ & $16.90 \pm 0.10$ & $24.69 \pm 0.02$ & 11 or 90\tablenotemark{c} \\
 
 ESO\,257-G?017    & J0727--45 & $07^{\rm h}\,27^{\rm m}$\,33\fs0 & --45\degr\,41\arcmin\,04\arcsec{} & 257\fdg8 & --13\fdg2 & $15.2 \pm 1.9$ & $1015 \pm 2$ & $ 730 \pm 2$ & $115 \pm 4$ & $0.62 \pm 0.10$ & $15.82 \pm 0.10$ & $25.35 \pm 0.03$ & 70 \\

 ESO\,368-G004     & J0732--35 & $07^{\rm h}\,32^{\rm m}$\,54\fs0 & --35\degr\,29\arcmin\,19\arcsec{} & 249\fdg0 &  --7\fdg7 & $14.5 \pm 1.6$ & $1383 \pm 2$ & $1105 \pm 2$ & $ 87 \pm 2$ & $1.94 \pm 0.31$ & $16.07 \pm 0.07$ & $24.72 \pm 0.02$ & 40 \\

 PGC\,023156       & J0815--28 & $08^{\rm h}\,15^{\rm m}$\,42\fs0 & --28\degr\,51\arcmin\,21\arcsec{} & 248\fdg0 &    3\fdg4 & $13.0 \pm 2.0$ & $1693 \pm 2$ & $1415 \pm 2$ & $163 \pm 6$ & $1.70 \pm 0.27$ & $16.91 \pm 0.10$ & $24.97 \pm 0.01$ & 70 \\

 ESO\,164-G?010    & J0826--54 & $08^{\rm h}\,26^{\rm m}$\,12\fs0 & --54\degr\,02\arcmin\,00\arcsec{} & 269\fdg9 &  --9\fdg2 & $11.2 \pm 1.8$ & $1052 \pm 2$ & $ 756 \pm 2$ & $158 \pm 3$ & $1.55 \pm 0.25$ & $13.91 \pm 0.07$ & $23.61 \pm 0.01$ & 35 \\

 \esoq{}           & J1057--48 & $10^{\rm h}\,57^{\rm m}$\,29\fs4 & --48\degr\,10\arcmin\,40\arcsec{} & 284\fdg1 &   10\fdg5 &  $122 \pm 4$   & $ 597 \pm 1$ & $ 311 \pm 1$ & $ 90 \pm 4$ & $0.95 \pm 0.15$ & $16.13 \pm 0.07$ & $25.48 \pm 0.02$ & 35 \\
 
 CGCG\,012-022     & J1133--03 & $11^{\rm h}\,33^{\rm m}$\,45\fs0 & --03\degr\,26\arcmin\,16\arcsec{} & 268\fdg4 &   54\fdg2 & $13.0 \pm 2.0$ & $1601 \pm 1$ & $1426 \pm 1$ & $141 \pm 2$ & $0.15 \pm 0.02$ & $15.06 \pm 0.02$ & $23.43 \pm 0.04$ & 50 \\

 UGC\,06780        & J1148--02 & $11^{\rm h}\,48^{\rm m}$\,50\fs0 & --02\degr\,01\arcmin\,56\arcsec{} & 273\fdg1 &   57\fdg2 & $21.0 \pm 3.0$ & $1723 \pm 2$ & $1561 \pm 2$ & $226 \pm 3$ & $0.09 \pm 0.01$ & $13.91 \pm 0.08$ & $24.19 \pm 0.05$ & 75 \\

 ESO\,572-G009     & J1153--18 & $11^{\rm h}\,53^{\rm m}$\,23\fs0 & --18\degr\,10\arcmin\,00\arcsec{} & 284\fdg1 &   42\fdg6 & $ 7.2 \pm 1.3$ & $1740 \pm 4$ & $1529 \pm 4$ & $ 49 \pm 2$ & $0.16 \pm 0.03$ & $16.79 \pm 0.05$ & $26.03 \pm 0.02$ & 50 \\
 
 ESO\,505-G007     & J1203--25 & $12^{\rm h}\,03^{\rm m}$\,31\fs0 & --25\degr\,28\arcmin\,36\arcsec{} & 289\fdg5 &   36\fdg2 &   $21 \pm 3$   & $1776 \pm 2$ & $1548 \pm 2$ & $ 88 \pm 5$ & $0.36 \pm 0.06$ & $14.20 \pm 0.06$ & $24.00 \pm 0.01$ & 50 \\
 
 ESO\,572-G052     & J1206--21 & $12^{\rm h}\,06^{\rm m}$\,03\fs0 & --21\degr\,11\arcmin\,54\arcsec{} & 289\fdg0 &   40\fdg5 & $ 6.0 \pm 1.3$ & $1791 \pm 2$ & $1575 \pm 2$ & $ 45 \pm 2$ & $0.26 \pm 0.04$ & $15.91 \pm 0.06$ & $23.72 \pm 0.03$ & 40 \\

 UGCA\,289         & J1235--07 & $12^{\rm h}\,35^{\rm m}$\,38\fs0 & --07\degr\,52\arcmin\,35\arcsec{} & 296\fdg1 &   54\fdg8 & $22.8 \pm 1.7$ & $ 980 \pm 2$ & $ 825 \pm 2$ & $172 \pm 2$ & $0.12 \pm 0.02$ & $13.83 \pm 0.03$ & $24.49 \pm 0.02$ & 55 \\

 UGCA\,307         & J1253--12 & $12^{\rm h}\,53^{\rm m}$\,57\fs0 & --12\degr\,06\arcmin\,30\arcsec{} & 303\fdg9 &   50\fdg8 & $22.8 \pm 1.3$ & $ 821 \pm 1$ & $ 664 \pm 1$ & $ 84 \pm 2$ & $0.24 \pm 0.04$ & $14.59 \pm 0.03$ & $24.25 \pm 0.03$ & 80 \\

 UGCA\,312         & J1259--12 & $12^{\rm h}\,59^{\rm m}$\,06\fs0 & --12\degr\,13\arcmin\,40\arcsec{} & 305\fdg9 &   50\fdg6 & $12.9 \pm 1.2$ & $1303 \pm 2$ & $1149 \pm 2$ & $102 \pm 2$ & $0.18 \pm 0.03$ & $15.16 \pm 0.02$ & $24.47 \pm 0.02$ & 45 \\

 UGCA\,322         & J1304--03 & $13^{\rm h}\,04^{\rm m}$\,30\fs0 & --03\degr\,34\arcmin\,22\arcsec{} & 309\fdg3 &   59\fdg1 & $38.0 \pm 2.0$ & $1357 \pm 2$ & $1238 \pm 2$ & $128 \pm 2$ & $0.13 \pm 0.02$ & $13.22 \pm 0.03$ & $24.09 \pm 0.01$ & 50 \\

 IC\,4212          & J1311--06 & $13^{\rm h}\,12^{\rm m}$\,03\fs0 & --06\degr\,59\arcmin\,29\arcsec{} & 312\fdg0 &   55\fdg5 & $46.0 \pm 1.0$ & $1476 \pm 1$ & $1350 \pm 1$ & $172 \pm 2$ & $0.19 \pm 0.03$ & $14.11 \pm 0.04$ & $24.42 \pm 0.02$ & 40 \\
 
 IC\,4824          & J1913--62 & $19^{\rm h}\,13^{\rm m}$\,14\fs0 & --62\degr\,05\arcmin\,18\arcsec{} & 334\fdg2 & --26\fdg3 & $18.0 \pm 1.2$ & $ 937 \pm 2$ & $ 820 \pm 2$ & $ 82 \pm 2$ & $0.24 \pm 0.04$ & $14.48 \pm 0.11$ & $24.19 \pm 0.04$ & 65 \\

 ESO\,141-G042     & J1916--62 & $19^{\rm h}\,16^{\rm m}$\,11\fs0 & --62\degr\,21\arcmin\,42\arcsec{} & 334\fdg0 & --26\fdg7 & $10.9 \pm 1.3$ & $ 896 \pm 2$ & $ 779 \pm 2$ & $126 \pm 4$ & $0.25 \pm 0.04$ & $13.70 \pm 0.09$ & $24.27 \pm 0.03$ & 90 \\

 IC\,4870          & J1937--65 & $19^{\rm h}\,37^{\rm m}$\,37\fs0 & --65\degr\,48\arcmin\,40\arcsec{} & 330\fdg3 & --29\fdg3 & $20.1 \pm 1.1$ & $ 875 \pm 1$ & $ 745 \pm 1$ & $ 96 \pm 2$ & $0.49 \pm 0.08$ & $14.79 \pm 0.06$ & $23.94 \pm 0.05$ & 65 \\

 IC\,4951          & J2009--61 & $20^{\rm h}\,09^{\rm m}$\,31\fs0 & --61\degr\,50\arcmin\,47\arcsec{} & 334\fdg9 & --32\fdg9 & $21.0 \pm 2.0$ & $ 800 \pm 2$ & $ 693 \pm 2$ & $127 \pm 4$ & $0.17 \pm 0.03$ & $14.09 \pm 0.02$ & $23.15 \pm 0.03$ & 85 \\

 ESO\,348-G009     & J2349--37 & $23^{\rm h}\,49^{\rm m}$\,23\fs5 & --37\degr\,46\arcmin\,19\arcsec{} & 349\fdg8 & --73\fdg2 & $13.1 \pm 0.3$ & $ 648 \pm 1$ & $ 633 \pm 1$ & $100 \pm 3$ & $0.06 \pm 0.01$ & $14.81 \pm 0.07$ & $24.79 \pm 0.03$ & 85 \\
 
 ESO\,149-G003     & J2352--52 & $23^{\rm h}\,52^{\rm m}$\,02\fs0 & --52\degr\,34\arcmin\,43\arcsec{} & 322\fdg5 & --62\fdg3 & $ 5.6 \pm 0.8$ & $ 590 \pm 2$ & $ 505 \pm 2$ & $ 74 \pm 4$ & $0.06 \pm 0.01$ & $14.79 \pm 0.03$ & $23.12 \pm 0.04$ & 90 \\

\enddata
\tablenotetext{a}{ESO\,121-G020 and \atg{} are unresloved in the HIPASS BGC (see \pII{})}
\tablenotetext{b}{Galaxy misidentified as CGMW\,1-0080 in HIPASS BGC}
\tablenotetext{c}{Inclination for ESO\,428-G033 highly uncertain due to star contamination (11\degr{} from \hi{} rotation curve, $\sim$90\degr{} from optical image).}

\begin{flushleft}
Notes.--- Col (1): Most commonly used galaxy name.  
Col (2): HIPASS source name.  
Col (3) \& (4): J2000.0 right ascension and declination as given in RC3 \citep[][ except \atg{}, which is taken from \pII{}]{dev91}.  
Col (5): Galactic longitude.  
Col (6): Galactic latitude.  
Col (7): Total integrated \hi{} flux density from the ATCA.  
Col (8): Systemic velocity from the \hi{} line (Heliocentric).  
Col (9): Velocity relative to the barycenter of the Local Group.  
Col (10): Velocity width of the \hi{} line at 20\% of the peak flux density.  
Col (11): SFD98 Galactic dust extinction value in the {\em B}-band.  
Col (12): Total apparent {\em B}-band magnitude from the 2.3m Telescope.  
Col (13): Effective {\em B}-band surface brightness, the average surface brightness out to the half light radius (Galactic extinction not applied).  
Col (14): Inclination estimated from the axis ratio of the 2.3m Telescope {\em B}-band images.
\end{flushleft}
\end{deluxetable}

\begin{deluxetable}{lcccccccc} 
\tabletypesize{\scriptsize}
\tablecaption{Summary of Derived Parameters for each Galaxy from ATCA and 2.3m Data.
\label{tab:summary}}
\tablewidth{0pt}
\tablehead{ \colhead{Name} & \colhead{$D$} & \colhead{\MB{}} & \colhead{$\log$\LB{}} & \colhead{$\log$\Mstar{}} & \colhead{$\log$\MHI{}} & \colhead{$\log$\Mbary{}} & \colhead{\mlr{}} & \colhead{\msmbr{}} \\

 \colhead{} & \colhead{(Mpc)} & \colhead{(mag)} & \colhead{($\log$\Lsun{})} & \colhead{($\log$\Msun{})} & \colhead{($\log$\Msun{})} & \colhead{($\log$\Msun{})} & \colhead{(\mmls{})} & \colhead{} \\
 
 \colhead{(1)} & \colhead{(2)} & \colhead{(3)} & \colhead{(4)} & \colhead{(5)} & \colhead{(6)} & \colhead{(7)} & \colhead{(8)} & \colhead{(9)}}
\startdata
 ESO\,349-G031  &  3.21\tablenotemark{a} & $-11.88 \pm 0.06$ & $6.94 \pm 0.06$ & $7.02 \pm 0.06$ &  $7.0 \pm 0.2$  & $7.4 \pm 0.2$ &  $1.2 \pm 0.3$  &  $0.4 \pm 0.1$  \\

 \mcg{}         &  9.5                   & $-14.65 \pm 0.06$ & $8.05 \pm 0.06$ & $8.13 \pm 0.06$ & $8.54 \pm 0.04$ & $8.8 \pm 0.1$ &  $3.0 \pm 0.3$  & $0.22 \pm 0.02$ \\

 ESO\,473-G024  &  7.6                   & $-13.11 \pm 0.06$ & $7.43 \pm 0.06$ & $7.51 \pm 0.06$ &  $7.9 \pm 0.2$  & $8.1 \pm 0.2$ &  $2.8 \pm 0.6$  & $0.23 \pm 0.05$ \\

 IC\,1574       &  4.92\tablenotemark{b} & $-13.62 \pm 0.03$ & $7.64 \pm 0.03$ & $7.72 \pm 0.03$ &  $7.5 \pm 0.2$  & $8.0 \pm 0.2$ & $0.64 \pm 0.13$ &  $0.6 \pm 0.1$  \\

 UGCA\,015      &  3.34\tablenotemark{b} & $-12.39 \pm 0.05$ & $7.15 \pm 0.05$ & $7.23 \pm 0.05$ &  $6.8 \pm 0.2$  & $7.4 \pm 0.2$ & $0.48 \pm 0.13$ &  $0.6 \pm 0.2$  \\

 ESO\,085-G047  & 16.3                   & $-16.99 \pm 0.04$ & $8.99 \pm 0.04$ & $9.07 \pm 0.04$ & $8.97 \pm 0.07$ & $9.4 \pm 0.1$ & $0.94 \pm 0.11$ & $0.48 \pm 0.06$ \\

 ESO\,120-G021  & 14.0                   & $-15.97 \pm 0.06$ & $8.58 \pm 0.05$ & $8.66 \pm 0.05$ &  $8.7 \pm 0.1$  & $9.1 \pm 0.1$ &  $1.3 \pm 0.2$  & $0.40 \pm 0.07$ \\

 ESO\,425-G001  & 14.9                   & $-16.07 \pm 0.06$ & $8.62 \pm 0.05$ & $8.70 \pm 0.05$ &  $8.5 \pm 0.3$  & $9.0 \pm 0.3$ &  $0.8 \pm 0.3$  &  $0.5 \pm 0.2$  \\

 UGCA\,120      &  8.6                   & $-17.03 \pm 0.07$ & $9.00 \pm 0.06$ & $9.08 \pm 0.06$ & $8.77 \pm 0.04$ & $9.3 \pm 0.1$ & $0.57 \pm 0.06$ & $0.60 \pm 0.06$ \\

 ESO\,121-G020  &  6.05\tablenotemark{a} & $-13.81 \pm 0.06$ & $7.72 \pm 0.05$ & $7.80 \pm 0.05$ & $7.90 \pm 0.03$ & $8.2 \pm 0.1$ & $1.49 \pm 0.13$ & $0.37 \pm 0.03$ \\
 
 \atg{}         &  6.05\tablenotemark{a} & $-12.07 \pm 0.07$ & $7.02 \pm 0.06$ & $7.10 \pm 0.06$ & $7.37 \pm 0.07$ & $7.7 \pm 0.1$ &  $2.2 \pm 0.3$  &  $0.3 \pm 0.3$  \\

 WHI\,B0619--07 &  7.7                   &  $-17.3 \pm 0.4$  &  $9.1 \pm 0.4$  &  $9.2 \pm 0.4$  & $8.86 \pm 0.06$ & $9.4 \pm 0.4$ &  $0.5 \pm 0.2$  &  $0.6 \pm 0.3$  \\

 ESO\,490-G017  &  4.23\tablenotemark{c} & $-15.49 \pm 0.07$ & $8.39 \pm 0.06$ & $8.47 \pm 0.06$ &  $7.5 \pm 0.1$  & $8.5 \pm 0.1$ & $0.13 \pm 0.03$ &  $0.9 \pm 0.2$  \\

 ESO\,255-G019  & 10.5                   & $-16.32 \pm 0.08$ & $8.72 \pm 0.07$ & $8.80 \pm 0.07$ & $8.86 \pm 0.07$ & $9.2 \pm 0.1$ & $1.37 \pm 0.19$ & $0.39 \pm 0.05$ \\

 ESO\,207-G007  & 10.7                   & $-16.77 \pm 0.06$ & $8.90 \pm 0.06$ & $8.98 \pm 0.06$ &  $8.6 \pm 0.1$  & $9.2 \pm 0.1$ & $0.55 \pm 0.09$ &  $0.6 \pm 0.1$  \\

 ESO\,207-G022  & 10.4                   & $-16.14 \pm 0.11$ &  $8.6 \pm 0.1$  &  $8.7 \pm 0.1$  &  $8.1 \pm 0.2$  & $8.9 \pm 0.2$ & $0.31 \pm 0.10$ &  $0.7 \pm 0.3$  \\

 ESO\,428-G033  & 19.5                   &  $-16.7 \pm 0.2$  &  $8.5 \pm 0.2$  &  $8.5 \pm 0.2$  & $9.11 \pm 0.02$ & $9.3 \pm 0.2$ &  $4.6 \pm 0.9$  & $0.16 \pm 0.03$ \\

 ESO\,257-G?017 &  9.7                   & $-14.75 \pm 0.14$ &  $8.1 \pm 0.1$  &  $8.2 \pm 0.1$  &  $8.5 \pm 0.1$  & $8.8 \pm 0.2$ &  $2.7 \pm 0.7$  & $0.24 \pm 0.06$ \\

 ESO\,368-G004  & 14.9                   &  $-16.7 \pm 0.3$  &  $8.9 \pm 0.3$  &  $9.0 \pm 0.3$  &  $8.9 \pm 0.1$  & $9.3 \pm 0.3$ &  $1.0 \pm 0.4$  &  $0.5 \pm 0.2$  \\

 PGC\,023156    & 19.0                   &  $-16.2 \pm 0.3$  &  $8.7 \pm 0.3$  &  $8.7 \pm 0.3$  &  $9.0 \pm 0.2$  & $9.3 \pm 0.3$ &  $2.4 \pm 1.0$  &  $0.3 \pm 0.1$  \\

 ESO\,164-G?010 & 10.1                   &  $-17.7 \pm 0.3$  &  $9.3 \pm 0.2$  &  $9.3 \pm 0.2$  &  $8.4 \pm 0.2$  & $9.4 \pm 0.3$ & $0.15 \pm 0.06$ &  $0.9 \pm 0.3$  \\

 \esoq{}        &  5.25\tablenotemark{d} &  $-13.4 \pm 0.2$  &  $7.6 \pm 0.2$  &  $7.6 \pm 0.2$  & $8.90 \pm 0.03$ & $9.1 \pm 0.2$ &   $22 \pm 4$    & $0.038 \pm 0.007$ \\

 CGCG\,012-022  & 19.1                   & $-16.49 \pm 0.03$ & $8.79 \pm 0.03$ & $8.87 \pm 0.03$ &  $9.0 \pm 0.2$  & $9.4 \pm 0.2$ &  $1.8 \pm 0.3$  & $0.32 \pm 0.06$ \\

 UGC\,06780     & 20.9                   & $-17.78 \pm 0.08$ & $9.30 \pm 0.07$ & $9.38 \pm 0.07$ &  $9.3 \pm 0.1$  & $9.7 \pm 0.1$ &  $1.1 \pm 0.2$  & $0.45 \pm 0.09$ \\

 ESO\,572-G009  & 20.4                   & $-14.92 \pm 0.06$ & $8.16 \pm 0.05$ & $8.24 \pm 0.05$ &  $8.9 \pm 0.2$  & $9.1 \pm 0.2$ &  $4.8 \pm 1.1$  & $0.15 \pm 0.03$ \\

 ESO\,505-G007  & 20.8                   & $-17.75 \pm 0.08$ & $9.29 \pm 0.08$ & $9.37 \pm 0.08$ & $9.37 \pm 0.03$ & $9.7 \pm 0.1$ & $1.18 \pm 0.12$ & $0.42 \pm 0.04$ \\

 ESO\,572-G052  & 21.0                   & $-15.96 \pm 0.07$ & $8.58 \pm 0.07$ & $8.66 \pm 0.07$ &  $8.8 \pm 0.2$  & $9.1 \pm 0.2$ &  $1.6 \pm 0.5$  & $0.34 \pm 0.09$ \\

 UGCA\,289      & 11.1                   & $-16.51 \pm 0.04$ & $8.80 \pm 0.03$ & $8.88 \pm 0.03$ & $8.82 \pm 0.07$ & $9.2 \pm 0.1$ & $1.04 \pm 0.11$ & $0.45 \pm 0.05$ \\

 UGCA\,307      &  8.9                   & $-15.39 \pm 0.05$ & $8.35 \pm 0.04$ & $8.43 \pm 0.04$ & $8.63 \pm 0.06$ & $8.9 \pm 0.1$ & $1.90 \pm 0.19$ & $0.31 \pm 0.03$ \\

 UGCA\,312      & 15.4                   & $-15.95 \pm 0.03$ & $8.57 \pm 0.03$ & $8.65 \pm 0.03$ & $8.86 \pm 0.09$ & $9.2 \pm 0.1$ &  $1.9 \pm 0.2$  & $0.31 \pm 0.04$ \\

 UGCA\,322      & 16.6                   & $-18.01 \pm 0.04$ & $9.40 \pm 0.03$ & $9.48 \pm 0.03$ & $9.39 \pm 0.05$ & $9.8 \pm 0.1$ & $0.98 \pm 0.08$ & $0.47 \pm 0.04$ \\

 IC\,4212       & 18.1                   & $-17.37 \pm 0.05$ & $9.14 \pm 0.05$ & $9.22 \pm 0.05$ & $9.55 \pm 0.02$ & $9.8 \pm 0.1$ & $2.56 \pm 0.17$ & $0.25 \pm 0.02$ \\

 IC\,4824       & 11.1                   & $-15.98 \pm 0.11$ & $8.58 \pm 0.10$ & $8.66 \pm 0.10$ & $8.72 \pm 0.07$ & $9.1 \pm 0.1$ &  $1.4 \pm 0.2$  & $0.39 \pm 0.07$ \\

 ESO\,141-G042  & 10.5                   & $-16.65 \pm 0.10$ & $8.85 \pm 0.09$ & $8.93 \pm 0.09$ &  $8.5 \pm 0.1$  & $9.1 \pm 0.1$ & $0.39 \pm 0.08$ &  $0.7 \pm 0.1$  \\

 IC\,4870       & 10.0                   & $-15.70 \pm 0.10$ & $8.47 \pm 0.09$ & $8.55 \pm 0.09$ & $8.68 \pm 0.05$ & $9.0 \pm 0.1$ &  $1.6 \pm 0.2$  & $0.35 \pm 0.05$ \\

 IC\,4951       &  9.4                   & $-15.95 \pm 0.03$ & $8.57 \pm 0.03$ & $8.65 \pm 0.03$ & $8.64 \pm 0.10$ & $9.0 \pm 0.1$ & $1.17 \pm 0.15$ & $0.42 \pm 0.05$ \\

 ESO\,348-G009  &  8.4                   & $-14.86 \pm 0.07$ & $8.14 \pm 0.07$ & $8.21 \pm 0.07$ & $8.34 \pm 0.02$ & $8.7 \pm 0.1$ & $1.58 \pm 0.14$ & $0.35 \pm 0.03$ \\

 ESO\,149-G003  &  6.5                   & $-14.33 \pm 0.03$ & $7.93 \pm 0.03$ & $8.00 \pm 0.03$ &  $7.7 \pm 0.1$  & $8.3 \pm 0.1$ & $0.66 \pm 0.11$ & $0.57 \pm 0.09$ \\
\enddata
\tablenotetext{a}{Distance for ESO\,349-G031 and ESO\,121-G020 taken from \citet{kar06} tip of the Red Giant Branch measurement.  \atg{} is assumed to be at the same distance as ESO\,121-G020.}
\tablenotetext{b}{Distance for IC\,1574 and UGCA\,015 taken from \citet{kar03b} tip of the Red Giant Branch measurement.}
\tablenotetext{c}{Distance for ESO\,490-G017 taken from \citet{kar03a} tip of the Red Giant Branch measurement.}
\tablenotetext{d}{Distance for \esoq{} distance taken from \citet{kar07} tip of the Red Giant Branch measurement.}

\begin{flushleft} %
Col (1) Galaxy name.  
Col (2) Distance estimate derived from the Local Group velocity (using H$_0$ = 75\kms{}\,Mpc$^{-1}$), except where a tip of the Red Giant Branch measurement was available (see notes).
Col (3) {\em B}-band absolute magnitude.  
Col (4) {\em B}-band luminosity.  
Col (5) Estimated stellar mass.  
Col (6) \hi{} mass.  
Col (7) Baryonic mass (stellar plus \hi{} mass).  
Col (8) \hi{} mass-to-{\em B}-band luminosity ratio (\mlr{}).  
Col (9) Stellar mass-to-baryonic mass ratio
\end{flushleft}
\end{deluxetable}

\begin{deluxetable}{lcccccccl} 
\tabletypesize{\scriptsize}
\tablecaption{Data for Other Galaxies Known to have High \hi{} Mass-to-Light Ratios.
\label{tab:other}}
\tablewidth{0pt}
\tablehead{ \colhead{Name} & \colhead{\MB{}} & \colhead{\LB{}} & \colhead{\MHI{}} & \colhead{\mlr{}} & \colhead{\AB{}, SFD98} & \colhead{D} & \colhead{\wxx{}/${\rm sin}i$} & \colhead{References} \\

   \colhead{}  & \colhead{(mag)} & \colhead{($\times 10^{7}$\Lsun{})} & \colhead{($\times 10^{7}$\Msun{})} & \colhead{(\mmls{})} & \colhead{(mag)} & \colhead{(Mpc)} & \colhead{(\kkms{})} & \colhead{} \\
 
 \colhead{(1)} & \colhead{(2)} & \colhead{(3)} & \colhead{(4)} & \colhead{(5)} & \colhead{(6)} & \colhead{(7)} & \colhead{(8)} & \colhead{(9)}}
\startdata
 DDO\,154   & $-13.8$ & $5.2$ & $49.0$ & $9.4$ & $0.04 \pm 0.01$ & $5.0$ & $156$ & \citet{car89} \\
    &  &  &  &  &  &  &  & \citet{hof93} \\
    &  &  &  &  &  &  &  & \citet{kru84} \\

 NGC\,3741  & $-13.1$ & $2.8$ & $16.0$ & $5.8$ & $0.11 \pm 0.02$ & $3.7$ & $119$ & \citet{beg05} \\
    &  &  &  &  &  &  &  & HyperLeda \\

 NGC\,2915  & $-15.9$ & $35.5$ & $95.8$ & $2.7$ & $1.19 \pm 0.19$ & $2.7$ & $188$ & \citet{meu94,meu96} \\

 UGCA\,292  & $-11.7$ & $0.72$ & $5.1$ & $7.0$ & $0.07 \pm 0.01$ & $4.3$ & $66$ & \citet{you03} \\
    &  &  &  &  &  &  &  & HyperLeda \\
\enddata

\begin{flushleft}
(1) Galaxy name. \\
(2) {\em B}-band absolute magnitude. \\
(3) {\em B}-band luminosity. \\
(4) \hi{} mass. \\
(5) \hi{} mass-to-light ratio. \\
(6) SFD98 {\em B}-band Galactic extinction value (\AB{}). \\
(7) Approximate distance derived from the Local Group velocity (using NED systemic velocity). \\
(8) Velocity width of the \hi{} line at 20\% of peak flux density corrected for inclination. \\
(9) Reference/s where we obtained the information on the galaxies.
\end{flushleft}
\end{deluxetable}

\begin{deluxetable}{llcclcc} 
\tabletypesize{\scriptsize}
\tablecaption{The Nearest Neighbors and Main Disturbers taken from NED for our Sample Galaxies and the Four Additional High \mlr{} Galaxies.
\label{tab:neigh}}
\tablewidth{0pt}
\tablehead{ \colhead{Name} & \multicolumn{2}{c}{Nearest Neighbor} & \colhead{ } & \multicolumn{3}{c}{Main Disturber Galaxy} \\ \cline{2-3} \cline{5-7}

\colhead{} & \colhead{Name} & \colhead{Separation} & & \colhead{Name} & \colhead{Separation} & \colhead{$\sim {\rm log}_{10}(L_{\rm B, MD})$} \\

\colhead{} & \colhead{} & \colhead{(Mpc)} & & \colhead{} & \colhead{(Mpc)} & \colhead{(${\rm log}_{10}$(\LLsun{}))} \\
 
 \colhead{(1)} & \colhead{(2)} & \colhead{(3)} & & \colhead{(4)} & \colhead{(5)} & \colhead{(6)}}
\startdata
 ESO\,349-G031  & NGC\,7793      & 0.34 & & NGC\,7793      & 0.34 & 9.28 \\

 \mcg{}         & NGC\,0024      & 1.62 & & NGC\,0024      & 1.62 & 9.35 \\

 ESO\,473-G024  & NGC\,0045      & 1.07 & & NGC\,0045      & 1.07 & 9.47 \\

 IC\,1574       & UGCA\,015      & 0.91 & & NGC\,0253      & 1.63 & 10.38 \\

 UGCA\,015      & NGC\,0253      & 0.79 & & NGC\,0253      & 0.79 & 10.22 \\

 ESO\,085-G047  & ESO\,085-G014  & 0.86 & & NGC\,1703      & 1.15 & 9.91 \\

 ESO\,120-G021  & ESO\,120-G012  & 0.74 & & ESO\,120-G006  & 1.34 & 10.37 \\

 ESO\,425-G001  & UGCA\,117      & 0.87 & & UGCA\,117      & 0.87 & 8.98 \\

 UGCA\,120      & ESO\,555-G028  & 0.48 & & ESO\,555-G028  & 0.48 & 7.36 \\

 ESO\,121-G020  & \atg{}         & 0.28 & & \atg{}         & 0.28 & 6.70 \\

 \atg{}         & ESO\,121-G020  & 0.28 & & ESO\,121-G020  & 0.28 & 7.38 \\

 WHI\,B0619--07 & UGCA\,127      & 0.28 & & UGCA\,127      & 0.28 & 9.31 \\

 ESO\,490-G017  & ESO\,489-G?056 & 0.23 & & ESO\,489-G?056 & 0.23 & 7.18 \\

 ESO\,255-G019  & ESO\,256-G013  & 0.71 & & ESO\,207-G007  & 0.91 & 8.91 \\

 ESO\,207-G007  & ESO\,207-G022  & 0.64 & & ESO\,207-G022  & 0.64 & 8.59 \\

 ESO\,207-G022  & ESO\,207-G007  & 0.64 & & ESO\,208-G021  & 0.79 & 9.67 \\

 ESO\,428-G033  & NGC\,2380      & 1.38 & & NGC\,2380      & 1.38 & 10.30 \\

 ESO\,257-G?017 & NGC\,2427      & 0.63 & & NGC\,2427      & 0.63 & 9.63 \\

 ESO\,368-G004  & \multicolumn{6}{c}{No Neighbors Identified, Deep in Galactic Plane}  \\

 PGC\,023156    & ESO\,431-G001  & 0.66 & & UGCA\,137      & 0.72 & 10.59 \\

 ESO\,164-G?010 & NGC\,2640      & 0.35 & & NGC\,2640      & 0.35 & 9.87 \\

 \esoq{}        & NGC\,4945      & 1.70 & & NGC\,4945      & 1.95 & 10.11 \\

 CGCG\,012-022  & CGCG\,012-003  & 0.35 & & CGCG\,012-003  & 0.35 & 8.90 \\

 UGC\,06780     & CGCG\,012-105  & 0.67 & & NGC\,3818      & 1.67 & 9.80 \\

 ESO\,572-G009  & ESO\,572-G006  & 0.15 & & ESO\,572-G006  & 0.15 & 8.32 \\

 ESO\,505-G007  & UGCA\,263      & 0.53 & & UGCA\,263      & 0.53 & 9.34 \\

 ESO\,572-G052  & UGCA\,257      & 0.80 & & UGCA\,257      & 0.80 & 9.51 \\

 UGCA\,289      & NGC\,4504      & 0.24 & & M\,104         & 0.88 & 10.75 \\

 UGCA\,307      & NGC\,4757      & 0.42 & & NGC\,4757      & 0.42 & 8.19 \\

 UGCA\,312      & NGC\,4920      & 0.47 & & NGC\,4781      & 0.85 & 9.97 \\

 UGCA\,322      & LCRS\,B130157.2--024313 & 0.88 & & UGC\,08041 & 1.27 & 9.65 \\

 IC\,4212       & NGC\,4948A     & 1.12 & & NGC\,4948A     & 1.12 & 8.99 \\

 IC\,4824       & AM\,1909--615  & 0.10 & & ESO\,141-G042  & 0.26 & 8.90 \\

 ESO\,141-G042  & AM\,1909--615  & 0.18 & & IC\,4824       & 0.26 & 8.54 \\

 IC\,4870       & NGC\,6744      & 0.80 & & NGC\,6744      & 0.80 & 10.75 \\

 IC\,4951       & ESO\,141-G042  & 1.95 & & ESO\,141-G042  & 1.95 & 8.73 \\

 ESO\,348-G009  & NGC\,7713      & 0.57 & & NGC\,7713      & 0.57 & 8.59 \\

 ESO\,149-G003  & ESO\,348-G009  & 1.97 & & ESO\,348-G009  & 1.97 & 7.93 \\
 
 \hline

 \multicolumn{7}{c}{Literature galaxies for comparison}  \\
 
 DDO\,154       & NGC\,4736      & 1.53 & & NGC\,4736      & 1.53 & 10.13 \\

 NGC\,3741      & LEDA\,166115   & 0.28 & & UGC\,06541     & 0.38 & 7.57 \\

 NGC\,2915      & HIPASS\,J0851--75 & 0.22 & & Circinus    & 1.35 & 10.15 \\

 UGCA\,292      & NGC4395        & 0.26 & & NGC4395        & 0.26 & 9.25 \\
\enddata

\begin{flushleft}
(1) Galaxy name. \\
(2) Name of the galaxy identified as the nearest neighbor. \\
(3) Approximate separation between the sample galaxy and the nearest neighbor in Mpc. \\
(4) Name of the galaxy identified as the main disturber. \\
(5) Approximate separation between the sample galaxy and the main disturber in Mpc. \\
(6) Approximate luminosity of the main disturber galaxy.
\end{flushleft}
\end{deluxetable}

\clearpage


\clearpage

\appendix

\section{Formulae}
\label{sec:equations}

The following is a list of formulae used to derive the parameters for our sample galaxies presented in Table~\ref{tab:summary}.

\begin{itemize}
\item The \hi{} mass \citep{rob75}

\begin{equation}
{\cal M}_{\rm HI} = 2.36 \times 10^{5}~D^{2}~F_{\rm HI} ~~{\cal M}_{\sun},
\end{equation}
where \FHI{} is the integrated \hi{} flux density in \jjks{}, and $D$ is the galaxy distance in Mpc.  Hubble-flow distances were calculated from the Local Group velocities given in the HIPASS BGC (using our adopted H$_0$ of 75\kms\,Mpc$^{-1}$).

\item The absolute {\em B}-band magnitude

\begin{equation}
M_{\rm B,0} = m_{\rm B} - A_{\rm B} - 5\log(D) - 25 ~~{\rm mag} ,
\end{equation}
where $D$ is the galaxy distance in Mpc, \mB{} is the {\em B}-band apparent magnitude, and \AB{} is the {\em B}-band Galactic extinction.

\item The {\em B}-band luminosity

\begin{equation}
L_{\rm B} = D^{2}~10^{10 - 0.4(m_{\rm B}-A_{\rm B}-M_{{\rm B}, \sun})} ~~L_{\sun,{\rm B}},
\end{equation} 
where $D$ is the galaxy distance in Mpc, \mB{} is the {\em B}-band apparent magnitude, \AB{} is the {\em B}-band Galactic extinction, and $M_{{\rm B}, \sun}$ is the absolute Solar {\em B} magnitude, which is taken as 5.48 mag \citep*{bes98}.

\item The \hi{} mass-to-light ratio

\begin{equation}
\frac{{\cal M}_{\rm HI}}{L_{\rm B}} = 1.5 \times 10^{-7} F_{\rm HI}~ 10^{0.4(m_{\rm B}-A_{\rm B})}~~\frac{{\cal M}_{\sun}}{L_{\sun,{\rm B}}} ,
\end{equation} 
where \FHI{} is the integrated \hi{} flux density in \jjks{}, \mB{} is the {\em B}-band apparent magnitude, and \AB{} is the {\em B}-band Galactic extinction.

\item Galactic extinction correction

\begin{equation}
A_{\rm B} = 4.32 E(B - V) ~~{\rm mag},
\end{equation}
where \AB{} is the {\em B}-band Galactic extinction corrections in magnitudes for their respective bands, and $E(B-V)$ is the difference between the extinction corrections in the {\em B} and {\em V}-bands.  (Based on the \citet*{sch98} extinctions as presented in NED\footnote{http://nedwww.ipac.caltech.edu/})

\item The Local Group velocity \citep{kar96a}

\begin{equation}
v_{\rm LG} = v_{\rm sys} + 300 \sin(l)\cos(b) ~~{\rm km\,s}^{-1},
\end{equation} 
where \vsys{} is the systemic velocity in \kkms{}, and $l$ and $b$ are the Galactic longitude and latitude, respectively, in degrees.

\end{itemize}


\clearpage


\begin{thebibliography}{}
\bibitem[Beaulieu et al.(2006)]{bea06} Beaulieu, S.F., Freeman, K.C., Carignan, C., Lockman, F.J., \& Jerjen, H., 2006, \aj, 131, 325
\bibitem[Begum et al.(2005)Begum, Chengalur, \& Karachentsev]{beg05} Begum, A., 
    Chengalur, J. N., \& Karachentsev, I. D., 2005, \aap, 433L, 1
\bibitem[Bekki et al.(2005)Bekki, Koribalski, \& Kilborn]{bek05} Bekki, K., Koribalski, B. S.,
    \& Kilborn, V. A., 2005, \mnras, tmpL, 77
\bibitem[Bell \& de Jong(2001)]{bel01} Bell, E. F., \& de Jong, R. S., 2001, \apj, 550, 212
\bibitem[Bessel et al.(1998)Bessell, Castelli, \& Plez]{bes98} Bessell, M. S.,
    Castelli, F., \& Plez, B., 1998, \aap, 333, 231
\bibitem[Binggeli et al.(1987)Binggeli, Tammann, \& Sandage]{bin87} Binggeli, B.,
    Tammann, G.A., \& Sandage, A., 1987, \aj, 94, 251
\bibitem[Binggeli et al.(1990)Binggeli, Tarenghi, \& Sandage]{bin90} Binggeli, B.,
    Tarenghi, M., \& Sandage, A., 1990, \aap, 228, 42
\bibitem[Blanton et al.(2005)]{bla05} Blanton, M. R., Eisenstein, D., Hogg, D. W., Schlegel, D. J., 
    \& Brinkmann, J., 2005, \apj{}, 629, 143
\bibitem[Bouchard et al.(2005)]{bou05} Bouchard, A., Jerjen, H., Da Costa, G. S., Ott, J.,
    2005, \aj, 130, 2058
\bibitem[Bouchard et al.(2007)]{bou07} Bouchard, A., Jerjen, H., Da Costa, G. S., Ott, J.,
    2007, \aj, 133, 261
\bibitem[Bothun et al.(1987)]{bot87} Bothun, G. D., Impey, C. D., Malin, D. F.,
    \& Mould, J. R., 1987, \aj, 94, 23
\bibitem[Broadhurst et al.(1992)Broadhurst, Ellis, \& Sandage]{bro92} Broadhurst, T. J., 
    Ellis, R. S., Glazebrook, K., 1992, Nature, 355, 55
\bibitem[Carignan \& Freeman(1988)]{car88} Carignan, C., \& Freeman, K. C., 1988, 
    \apj, 332, L33
\bibitem[Carignan \& Beaulieu(1989)]{car89} Carignan, C., \& Beaulieu, S., 1989, 
    \apj, 347, 760
\bibitem[Carignan \& Purton(1998)]{car98} Carignan, C., \& Purton, C., 1998, 
    \apj, 506, 125
\bibitem[Davies et al.(2006)]{dav06} Davies, J. I., Disney, M. J., Minchin, R. F., 
    Auld, R., \& Smith, R., 2006, \mnras, 368, 1479
\bibitem[de Blok et al.(1996)de Blok, McGaugh, \& van der Hulst]{deb96} 
    de Blok, W. J. G., McGaugh, S. S., \& van der Hulst, J. M., 1996, \mnras, 283, 18
\bibitem[De Rijcke et al.(2007)]{der07} De Rijcke, S., Zeilinger, W. W., Hau, G. K. T.,
    Dejonghe, H., \& Prugniel, P., 2007, \apj, in press, astro-ph/0701424
\bibitem[de Vaucouleurs et al.(1991)]{dev91} de Vaucouleurs, G., de
    Vaucouleurs, A., Corwin, H. G., Buta, R. J., Paturel, G., \& Fouqu\'{e}, 
    P., 1991, Third Reference Catalogue of Bright Galaxies, Springer, New York
\bibitem[Disney(1976)]{dis76} Disney, M. J., 1976, Nature, 263, 573
\bibitem[Doyle et al.(2005)]{doy05} Doyle, M. T., et al., 2005,\mnras, 361, 34
\bibitem[Dressler(1980)]{dre80} Dressler, A., 1980, \apj, 236, 351
\bibitem[Dressler(1984)]{dre84} Dressler, A., 1984, \araa, 22, 185
\bibitem[Geha et al.(2006)]{geh06} Geha, M., Blanton, M. R., Masjedi, M., 
    \& West, A. A., 2006, \apj, 653, 240 
\bibitem[Gentile et al.(2007)]{gen07} Gentile, G., Salucci, P., Klein, U., 
    \& Granato, G., 2007, \mnras, 375, 199 
\bibitem[Grebel, \& Gallagher(2004)]{gre04} Grebel, E. K., \& Gallagher, J. S., III, 2004, \apj, 610, L89
\bibitem[Grossi et al.(2007)]{gro07} Grossi, M., Disney, M. J., Pritzl, B. J., 
    Knezek, P. M., Gallagher, J. S., Minchin, R. F., \& Freeman, K. C., 2007, 
    \mnras, 374, 107 
\bibitem[Gurovich et al.(2004)]{gur04} Gurovich, S., McGaugh, S. S., Freeman, K. C., 
    Jerjen, H., Staveley-Smith, L., \& de Blok, W. J. G., 2004, PASA, 21, 412
\bibitem[Haynes et al.(1999)]{hay99} Haynes, M. P., Giovanelli, R., Chamaraux, P., 
    da Costa, L. N., Freudling, W., Salzer, J. J., \& Wegner, G., 1999, \aj, 117, 2039
\bibitem[Hoffman et al.(1993)]{hof93} Hoffman, G. L., Lu, N. Y.,
    Salpeter, E. E., Farhat, B., Lamphier, C., \& Roos, T., 1993, \aj, 106, 39
\bibitem[Ibata et al.(1994)Ibata, Gilmore, \& Irwin]{iba94} Ibata, R. A., 
    Gilmore, G., \& Irwin, M. J., 1994, Nature, 370, 194
\bibitem[Impey \& Bothun (1989)]{imp89} Impey, C. D., \& Bothun, G. D., 
    1989, \apj, 341, 89
\bibitem[Jerjen et al.(2000)Jerjen, Binggeli, \& Freeman]{jer00} Jerjen, H., Binggeli, B., \& Freeman, K.C., 2000, \aj, 119, 593
\bibitem[Jerjen(2003)]{jer03} Jerjen, H., 2003, \aap, 398, 63
\bibitem[Karachentsev \& Makarov(1996)]{kar96a} Karachentsev, I. D., \& Makarov,
    D. I., 1996, \aj, 111, 794
\bibitem[Karachentsev \& Makarov(1999)]{kar99} Karachentsev, I. D., \& Makarov,
    D. I.,  1999, IAU Symp. 186, Galaxy Interactions at High and Low Redshift,
    ed J. Barnes \& D. B. Sanders (Doredrecht: Klewer), 109
\bibitem[Karachentsev et al.(2003a)]{kar03a} Karachentsev, I. D., et al., 2003a, \aap, 398, 479
\bibitem[Karachentsev et al.(2003b)]{kar03b} Karachentsev, I. D., et al., 2003b, \aap, 404, 93
\bibitem[Karachentsev et al.(2004)]{kar04} Karachentsev, I. D., Karachentseva, V. E.,
    Huchtmeier, W. K., \& Makarov, D. I., 2004, \aj, 127, 2031
\bibitem[Karachentsev et al.(2006)]{kar06} Karachentsev, I. D., et al., 2006, \aj, 131, 1361
\bibitem[Karachentsev et al.(2007)]{kar07} Karachentsev, I. D., et al., 2007, \aj, 133, 504
\bibitem[Kennicutt \& Skillman(2001)]{ken01} Kennicutt, R. C., \& Skillman, E. D., 
    2001, \aj, 121, 1461
\bibitem[Koribalski et al.(2004)]{kor04} Koribalski, B. S., et al., 2004, 
    \aj, 128, 16
\bibitem[Krumm \& Burstein(1984)]{kru84} Krumm, N., \& Burstein, D., 1984, \aj, 89, 1319
\bibitem[Landolt(1992)]{lan92} Landolt, A. U., 1992, \aj, 104, 340
\bibitem[Leroy et al.(2005)]{ler05} Leroy, A., Bolatto, A. D., Simon, J. D., \& Blitz, L.,
    2005, \apj, 625, 763
\bibitem[Mateo(1998)]{mat98} Mateo, M., 1998, \araa, 36, 435
\bibitem[Matthews et al.(1998)Matthews, van Driel, \& Gallagher]{matt98} Matthews, L. D., 
    van Driel, W., \& Gallagher, J. S. 1998, \aj, 116, 1169
\bibitem[Matthews et al.(2001)Matthews, van Driel, \& Monnier-Ragaigne]{mat01} Matthews, 
    L. D., van Driel, W., \& Monnier-Ragaigne, D., 2001, \aap, 365, 1
\bibitem[McGaugh et al.(2000)]{mcg00} McGaugh, S. S., Schombert, J. M., Bothun,
    G. D., \& de Blok, W. J. G., 2000, \apj, 533L, 99
\bibitem[McGaugh(2005)]{mcg05} McGaugh, S. S., 2005, \apj, 632, 859
\bibitem[Meurer et al.(1994)Meurer, Mackie, \& Carignan]{meu94} 
    Meurer, G. R., Mackie, G., \& Carignan, C., 1994, \aj, 107, 2021
\bibitem[Meurer et al.(1996)]{meu96} Meurer, G. R., Carignan, C.,
    Beaulieu, S. F., \& Freeman, F. C., 1996, \aj, 111, 155
\bibitem[Mo et al.(1998)Mo, Mao, \& White]{mo98} Mo, H. J., Mao, S., \& White, S. D. M.,
    1998, \mnras, 295, 319
\bibitem[Oemler(1974)]{oem74} Oemler, A. Jr., 1974, \apj, 194, 1
\bibitem[Paturel et al.(2003)]{pat03} Paturel, G., Petit, C., Prugniel, P., Theureau, G.,
    Rousseau, J., Brouty, M., Dubois, P., \& Cambrésy, L., 2003, \aap, 412, 45,
    http://leda.univ-lyon1.fr/
\bibitem[Perryman et al.(2001)]{per01} Perryman, M. A. C., et al., 2001, \aap, 369, 339, http://www.rssd.esa.int/gaia/
\bibitem[Pfenniger \& Revaz(2005)]{pfe05} Pfenniger, D., \& Revaz, Y., 2005, \aap, 431, 511
\bibitem[Pickering et al.(1997)]{pic97} Pickering, T. E., Impey, C. D., van Gorkom, J. H., 
    \& Bothun, G. D., 1997, \aj, 114, 1858
\bibitem[Pierce \& Tully(1988)]{pie88} Pierce, M. J., \& Tully, R. B., 1988, \apj, 330, 579
\bibitem[Postman \& Geller(1984)]{pos84} Postman, M., \& Geller, M. J., 1984, \apj, 281, 95
\bibitem[Rejkuba et al.(2006)]{rej06} Rejkuba, M., Da Costa, G.S., Jerjen, H., Zoccali, M., \& Binggeli, B., 2006, \aap, 448, 983
\bibitem[Rhee(2004)]{rhe04} Rhee, M.-H., 2004, JKAS, 37, 91
\bibitem[Rix, \& Rieke(1993)]{rix93} Rix, H.-W., \& Rieke, M. J., 1993, \apj, 418,123
\bibitem[Roberts(1975)]{rob75} Roberts, M. S., 1975, Galaxies and the Universe, 
    University of Chicago Press, Chicago
\bibitem[Roberts \& Haynes(1994)]{rob94} Roberts, M. S., \& Haynes,
    M. P., 1994, \araa, 32, 115
\bibitem[Ryan-Weber et al.(2002)]{rya02} Ryan-Weber, E., et al., 2002, \aj, 124, 1954
\bibitem[Salpeter(1955)]{sal55} Salpeter, E. E., 1955, \apj, 121, 161
\bibitem[Sauty et al.(2003)]{sau03} Sauty, S., 2003, \aap, 411, 381
\bibitem[Schlegel et al.(1998)Schlegel, Finkbeiner, \& Davis]{sch98} 
    Schlegel, D. J., Finkbeiner, D. P., \& Davis, M., 1998, \apj, 500, 525
\bibitem[Schombert et al.(1992)]{sch92} Schombert, J. M., Bothun, G. D., 
    Schneider, S. E., \& McGaugh, S. S., 1992, \aj, 103, 1107
\bibitem[Skillman et al.(2003)Skillman, C\^ot\'e, \& Miller]{ski03b} 
    Skillman, E. D., C\^ot\'e, S., \& Miller, B. W., 2003, \aj, 125, 610
\bibitem[Steinmetz et al.(2006)]{ste06} Steinmetz, M., et al., 2006, \aj, 132, 1645, http://www.rave-survey.org
\bibitem[Taylor et al.(1998)]{tay98} Taylor, C. L., Kobulnicky, H. A., 
    \& Skillman, E. D., 1998, \aj, 116, 2746
\bibitem[Taylor \& Webster(2005)]{tay05} Taylor, E. N., \& Webster, R. L.,
    2005, \apj, 634, 1067
\bibitem[Tolstoy(1999)]{tol99} Tolstoy, E., 1999, \apss, 265, 199
\bibitem[Toomre(1964)]{too64} Toomre, A., 1964, \apj, 139, 1217
\bibitem[Tully \& Fisher(1977)]{tul77} Tully, R. B., \& Fisher, J. R., 1977, \aap, 54, 661
\bibitem[van Zee(2001)]{vzee01} van Zee, L., 2001, ApJ, 543, 31L
\bibitem[Verde et al.(2002)Verde, Oh, \& Jimenez]{ver02} 
    Verde, L., Oh, S. P., \& Jimenez, R., 2002, \mnras, 336, 541
\bibitem[Verheijen(1997)]{ver97} Verheijen, M. A. W., 1997, The Ursa Major Cluster of Galaxies: TF-relations and dark matter, Ph.D. Thesis, Univ. Groningen, The Netherlands, chaps. 5 and 6
\bibitem[Verheijen(2001)]{ver01} Verheijen, M. A. W., 2001, \apj, 563, 694
\bibitem[Warren et al.(2004)Warren, Jerjen, \& Koribalski]{war04} 
    Warren, B. E., Jerjen, H., \& Koribalski, B. S., 2004, \aj, 128, 1152
\bibitem[Warren(2005)]{war05} Warren, B. E., 2005, The Nature of High \hi{} Mass-to-Light Ratio Field Galaxies, Ph.D. Thesis, Australian National Univ., Australia
\bibitem[Warren et al.(2006)Warren, Jerjen, \& Koribalski]{war06} 
    Warren, B. E., Jerjen, H., \& Koribalski, B. S., 2006, \aj, 131, 2056
\bibitem[Williams et al.(1996)]{wil96} Williams, R. E., et al., 1996, \aj, 112, 1335
\bibitem[York et al.(2000)]{yor00} York, D. G., 2000, \aj, 120, 1579, http://www.sdss.org/
\bibitem[Young et al.(2003)]{you03} Young, L. M., van Zee, L., Lo, K. Y., 
    Dohm-Palmer, R. C., Beierle, M. E., 2003, \apj, 592, 111

\end{thebibliography}
\end{document}